\documentclass[sigconf]{acmart}
\copyrightyear{2023}
\acmYear{2023}
\setcopyright{rightsretained}
\acmConference[CAIN '23]{2023 IEEE/ACM 2nd International Conference on AI Engineering – Software Engineering for AI}{May 15–16, 2023}{Melbourne, Australia}
\acmBooktitle{2023 IEEE/ACM 2nd International Conference on AI Engineering – Software Engineering for AI (CAIN '23), May 15–16, 2023, Melbourne, Australia}

\usepackage[T1]{fontenc}
\usepackage{paralist}
\usepackage{newunicodechar}
\usepackage{wrapfig}
\usepackage{tabularx}
\usepackage{tablefootnote}
\usepackage{multirow}
\usepackage{colortbl}
\usepackage{hyperref}
\usepackage{balance}
\usepackage{enumitem}

\newcommand{\myparagraph}[1]{\vspace{.5em}\noindent\textbf{#1.}\ }
\renewcommand\subsubsection[1]{\myparagraph{#1}}

\newcommand\tableSummary[0]{\begin{table}[t]
    \label{challengessummary}
    \small
\begin{tabularx}{\linewidth}{>{\columncolor[gray]{.9}}p{.98\linewidth}}
\textbf{Requirements Engineering:} 
Lack of AI literacy causes unrealistic expectations from customers, managers, and even other team members  $\bullet$ 
Vagueness in ML problem specifications makes it difficult to map business goals to performance metrics $\bullet$ 
Regulatory constraints specific to data and ML introduce additional requirements that restrict development \\
\textbf{Architecture, Design, and Implementation:} 
Transitioning from a model-centric to a pipeline-driven or system-wide view is considered important for moving into production, but a difficult paradigm shift for many teams $\bullet$ 
ML adds substantial design complexity with many, often implicit, data and tooling dependencies, and entanglements due to a lack of modularity $\bullet$ 
Difficulty in scaling model training and deployment on diverse hardware $\bullet$ 
While monitorability and planning for change are often considered important, they are mostly considered only late after launching \\
\textbf{Model Development:}
Model development benefits from engineering infrastructure and tooling but provided infrastructure and technical support are limited in many teams $\bullet$ 
Code quality is not standardized in model development tools, leading to conflicts about code quality \\
\textbf{Data Engineering:}
Data quality is considered important, but difficult for practitioners and not well supported by tools $\bullet$ 
Internal data security and privacy policies restrict data access and use  $\bullet$ 
Although training-serving skew is common, many teams lack support for its required detection and monitoring $\bullet$ 
Data versioning and provenance tracking are often seen as elusive, with not enough tool support \\
\textbf{Quality Assurance:}
Testing and debugging ML models is difficult due to lack of specifications $\bullet$ 
Testing of model interactions, pipelines, and the entire system is considered challenging and often neglected $\bullet$ 
Testing and monitoring models in production are considered important but difficult, and often not done $\bullet$ 
There are no standard processes or guidelines on how to assess system qualities such as  fairness, security, and safety in practice \\
\textbf{Process:}
Development of products with ML component(s) is often  ad-hoc, lacking well-defined processes  $\bullet$ 
The uncertainty in ML development makes it hard to plan and estimate effort and time \\
\textbf{Organization and Teams:}
Building products with ML components requires diverse skill sets, which is often missing in development teams $\bullet$ 
Many teams are not well prepared for the extensive interdisciplinary collaboration and communication needed in ML products $\bullet$ 
ML development can be costly and resource limits can substantially curb/limit efforts $\bullet$ 
Lack of organizational incentives, resources, and education hampers achieving all system-level qualities\\
\end{tabularx}
    \caption{Overview of Identified Challenges}
\end{table}}

\newcommand\tableA[0]{\begin{table}[t]
    \caption{Paper Selection}
    \label{paperselection}
    \small
\begin{tabularx}{\linewidth}{lrrr}
\toprule
\textbf{Data Source}                              & \multicolumn{1}{p{1cm}}{\textbf{Initial Search Result }} & \multicolumn{1}{p{2.5cm}}{\textbf{After Filtering by Title/Abstract and Snowballing}} & \multicolumn{1}{p{1.5cm}}{\textbf{Final Selection}} \\\midrule
IEEE             & 69                                   & 30          & 19          \\
ACM              & 48                                   & 11          & 10          \\
Willey           & 6                                    & 0           & 0           \\
ScienceDirect    & 32                                   & 5           & 3           \\
Engineer Village & 101                                  & 3           & 0           \\
Springer         & 6* & 3           & 2           \\
arXiv            & 79                                   & 8           & 5           \\
Snowballing      & -                                    & 26          & 11          \\
\textbf{Total} & \textbf{341}                     & \textbf{86}      & \textbf{50}
\\\bottomrule
\multicolumn{4}{l}{*abstract filtering from 5612 papers retrieved with fulltext search}                                                             \\
\end{tabularx}
\end{table}}

\newcommand\tableB[0]{
\begin{table}[t]
    \caption{Inclusion and Exclusion Criteria}
    \label{criteria}
    \small
\begin{tabularx}{\linewidth}{lX}
\toprule
\multicolumn{2}{l}{\textbf{Inclusion Criteria}}                                                             \\\midrule
I1: & Paper includes software engineering challenges for ML systems                                \\
I2: & Paper uses interview or survey with industry practitioners (software engineers, data scientists, etc.) to identify the challenges           \\
I3: & Paper appears in a refereed publication (including conference proceedings, journal, etc.) or uploaded in arxiv in a publication format      \\
I4: & Paper is written in English                                                                  \\
\multicolumn{2}{l}{}                                                                               \\\midrule
\multicolumn{2}{l}{\textbf{Exclusion Criteria}}                                                            \\\midrule 
E1: & Paper has a strict  ML model view and does not consider the system or product using the model \\
E2: & Paper interviews/surveys only non-technical people (end-users, domain experts, etc.)        \\
E3: & Paper focuses on ML for software engineering instead of software engineering for ML systems  \\
E4: & Paper falls in the category of gray literature: blog post, technical report, government report, webinar, poster session, presentation, etc.\\\bottomrule
\end{tabularx}
\end{table}
}

\newcommand\figureA[0]{
\begin{figure}[t]
\includegraphics[width=.88\linewidth]{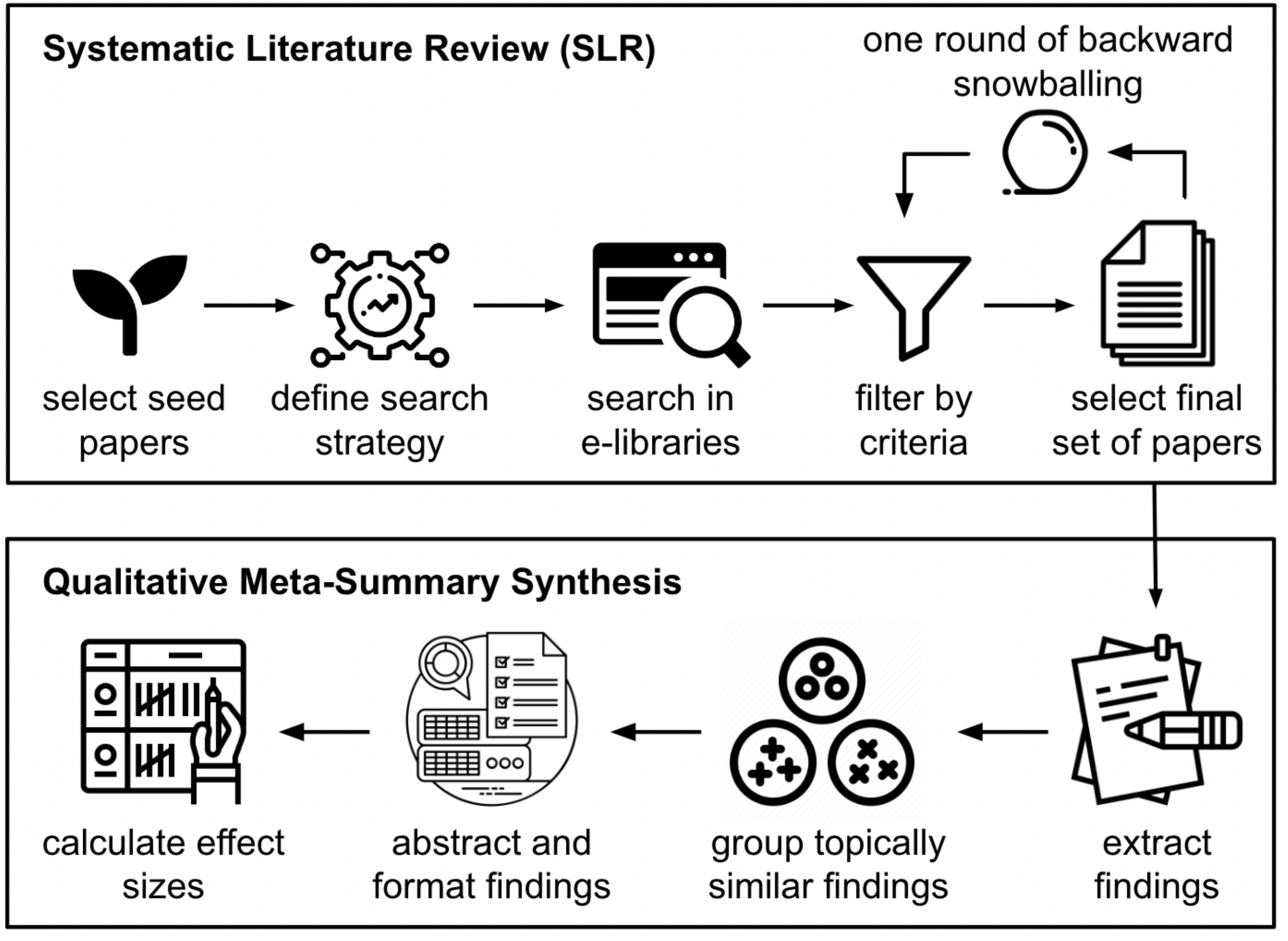}
\caption{Research Method}
\label{fig:overview}
\end{figure}
}

\newcommand\figureB[0]{
\begin{figure}[t]
\includegraphics[width=.6\linewidth]{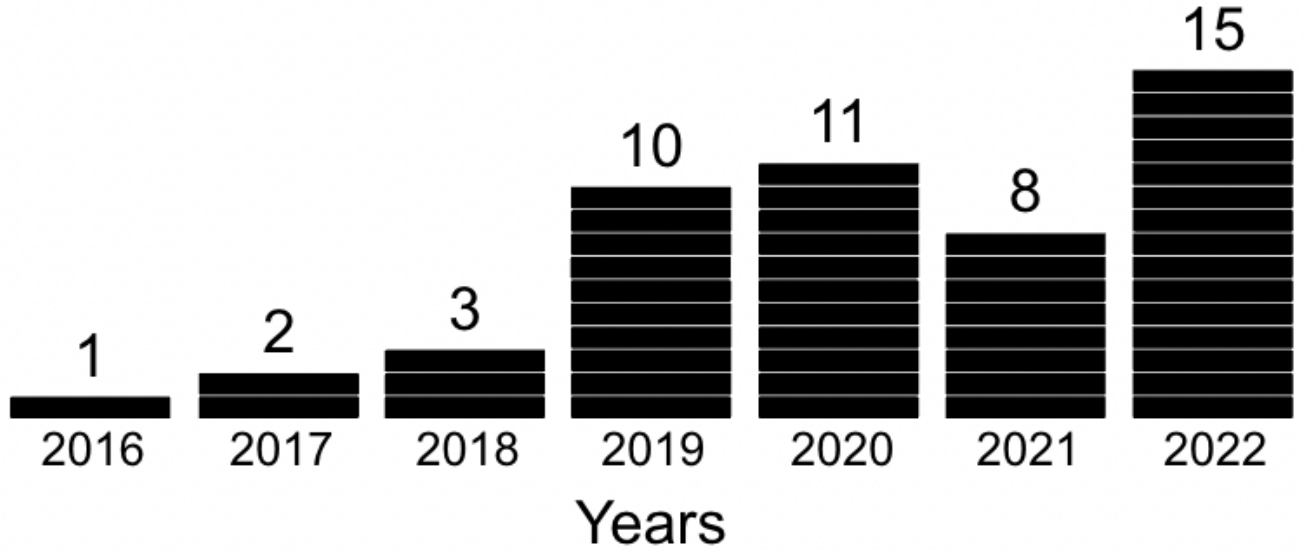}
\caption{Year Distribution of the Selected Papers}
\label{fig:yeardist}
\end{figure}
}

\renewcommand\footnotetextcopyrightpermission[1]{}

\begin{document}
\title{A Meta-Summary of Challenges in Building Products with ML Components -- Collecting Experiences from 4758+ Practitioners}
\author{Nadia Nahar}
\email{nadian@andrew.cmu.edu}
\affiliation{%
   \institution{Carnegie Mellon University}
   \city{Pittsburgh}
   \state{PA}
   \country{USA}}
\author{Haoran Zhang}
\affiliation{%
   \institution{Carnegie Mellon University}
   \city{Pittsburgh}
   \state{PA}
   \country{USA}}
\author{Grace Lewis}
\affiliation{%
   \institution{Carnegie Mellon Software Engineering Institute}
   \city{Pittsburgh}
   \state{PA}
   \country{USA}}
\author{Shurui Zhou}
\affiliation{%
   \institution{University of Toronto}
   \city{Toronto}
   \state{Ontario}
   \country{Canada}}
\author{Christian Kästner}
\affiliation{%
   \institution{Carnegie Mellon University}
   \city{Pittsburgh}
   \state{PA}
   \country{USA}}

\begin{abstract}
        Incorporating machine learning (ML) components into software products raises new software-engineering challenges and exacerbates existing challenges. Many researchers have invested significant effort in understanding the challenges of industry practitioners working on building products with ML components, through interviews and surveys with practitioners. With the intention to aggregate and present their collective findings, we conduct a meta-summary study: We collect 50 relevant papers that together interacted with over 4758 practitioners using guidelines for systematic literature reviews. We then collected, grouped, and organized the over 500 mentions of challenges within those papers. We highlight the most commonly reported challenges and hope this meta-summary will be a useful resource for the research community to prioritize research and education in this field.
\end{abstract}
                \maketitle

        \section{Introduction}\label{h.afl3gubgs1jy}
\tableSummary

After decades of effort in machine learning (ML) research to build better models, researchers from industry and academia have recently started to shift their attention to improving how to build software products with such models. Incorporating a ML component into a software product is often argued to be harder than incorporating traditional functional components, because of the specific characteristics of machine learning (e.g., based on data, no specifications in the traditional sense, fairness concerns) and how they impact the entire life cycle of the product \cite{5GnQ,xuax,8qrd,ZAdD,y46l}. While the traditional software development process has challenges of its own, bringing ML into the picture is argued to break a lot of existing software architecture and engineering assumptions \cite{5GnQ,y46l}. This leads initiatives to rethink existing processes and practices and shift priorities in software teams. As a result, we keep hearing from practitioners on how they perceive building, deploying, and incorporating machine learning in software products as a challenge, even when the initial ML research and model prototypes seemed promising. 

While some practitioners give talks on challenges or write experience papers (e.g., examples in academic venues surveyed elsewhere \cite{G64S,jAIR}), researchers have also been actively studying the challenges faced by practitioners when building software products with ML components across many projects. In recent years, many researchers have \emph{interviewed} or \emph{surveyed} practitioners to identify what has really changed for them with the introduction of machine learning, often with the goal of identifying challenges, research opportunities, and best practices in a rapidly changing field. While some studies focus on specific aspects, such as challenges regarding architecture \cite{Dk4s}, collaboration \cite{xMO3}, or fairness \cite{O0Hy,aqp3}, many others explore challenges more broadly. Many of these studies have identified similar challenges. We believe that we have reached a point where practices have settled and research on challenges approaches saturation -- we think that now is a good time to step back and survey the collective findings of the research community.

\figureA

In this paper, we aim to consolidate knowledge about challenges in the practice of building software products with ML components, with a systematic literature survey of existing studies that \emph{interviewed} or \emph{surveyed} industry practitioners across multiple projects. We identified 50 studies of which, 30 conducted interviews, 11 conducted surveys, and nine did both, with a total of over 4758 identified participants (seven studies did not report the number of participants; some participants may have participated in multiple studies). Using the \emph{meta-summary research method} \cite{ZbHh,J0kZ,UxiN}, we analyze, organize, and synthesize findings across all these studies (as shown in Figure \ref{fig:overview}), answering the overall research question: \textbf{What are the challenges experienced by industry practitioners in building software products with ML components?}

We group the challenges found in the meta-summary into categories. In a nutshell, we find practitioners struggle in different product development stages: (1) requirements engineering, (2) architecture, design, and implementation, and (3) quality assurance. We also find several engineering challenges in ML-specific stages, in particular (4) model development, and (5) data engineering. Other issues relate to cross-cutting concerns related to (6) process, and (7) organization and teams. Following the meta-summary method, we present and organize the challenges mentioned by the practitioners in the original papers as they have been reported, without attempting to speculate or pass our own judgments on the findings. Table \ref{challengessummary} contains a summary of the findings. We conclude the paper with a brief discussion, reflecting our own views.

Our key contribution is the meta-summary, which we present in narrative form in this paper. We additionally provide details with clear traceability to findings and papers as supplementary documents \cite{1JnV}.

\section{Scoping and Related Work}\label{h.xwve7xaxsvpt}
With the advance of ML techniques, many organizations have invested substantial efforts in building products with ML components. While there is a large amount of research that focuses entirely on the challenges that data scientists face in their model-development work (e.g., development responsibilities \cite{TtFP}, data exploration \cite{jM0j,clxs}, data-science processes \cite{XpO1,E5GQ}, development in notebooks \cite{66i9,rQUY,eI6i}, AutoML \cite{agH0}), another body of work focuses on the challenges of building products with those models, often with interdisciplinary teams, and placing substantial attention on qualities like safety and observability. The latter work, which forms the scope of this survey, moves beyond the model-centric view of classic data-science workflows and considers building automated pipelines and entire software systems with many ML and non-ML components, as well as the engineering challenges involved. It emerges in a growing research community often named \emph{Software Engineering for Machine Learning (SE4ML)} studying the engineering challenges of building both ML components and products that contain ML components.

\subsubsection{Understanding practitioner needs}\label{h.nzqztijxpr6o}
Academic research is often criticized for being far removed from the needs faced by practitioners in industry \cite{75w3,aqp3,bJXS}. If researchers want to achieve rapid impact in industry, they need to understand what problems are important to practitioners; conversely, practitioners may attempt to attract researchers to work on their problems. Attempts to close the gap between academia and practice typically need to navigate a tradeoff between (a) investigating one or few teams in depth with findings that may not generalize or (b) exploring common problems across many teams with more shallow engagements. Focused on individual teams, we see a few ethnographic studies \cite{5ezP,auDR}, many direct collaborations with an industrial partner \cite{kGqp,JvNS}, and many experience reports published by practitioners in papers \cite{x2RD,TcaQ,cHAy,JF8I}, talks \cite{uJDv,Ak3f,gVYg}, or blog posts \cite{0BTr,rN8P,Brt4}. To understand problems across teams, many researchers conduct interviews across multiple teams and organizations, e.g., \cite{DNDN,5GnQ,Dk4s,xMO3,x37l}, to be either addressed by the same researchers or reported as open problems to the community. Other researchers have focused on surveying practitioners at scale across companies and regions, e.g., \cite{R4n4,QNQY,IrZw,CN5W}. 

In this paper, we go one step further in aggregating and analyzing results from prior interviews and surveys with over 4758 practitioners, which we hope will help guide future research and educational activities toward challenges relevant to practitioners.

\subsubsection{Previous literature reviews}\label{h.2pg9hzdorkck}
There have been several prior literature reviews on topics related to building products with ML components. Most surveys review academic papers proposing solutions in subfields, such as testing ML components \cite{PSLS,rPJR,InOi,bQQ0,hW2b}, safety and security \cite{9HsR,LOJN,hW2b}, data management \cite{M56h}, and even trying to cover published research on SE4ML broadly \cite{1nxv,Jbnm}. The closest to our work are two literature surveys that analyze practitioner experience reports published at academic conferences (not including grey literature) collecting the self-reported challenges of a few dozen teams \cite{G64S,jAIR}. In this work, we specifically perform a meta-summary of academic papers reporting on interviews and surveys with practitioners.

\section{Research Method}\label{h.8hsm0hxan277}
The goal of this paper is to summarize challenges in building products with ML components, accumulated from industry practitioners in prior research. To achieve this goal and answer our research question, we first define the appropriate search strategy and study selection criteria to find the relevant literature that identifies challenges through communicating with industry practitioners. We follow the guidelines for systematic literature review (SLR) for this paper collection step \cite{XMGU}. Then, we extract the data from the selected papers and analyze the data to complete the synthesis process. Several approaches have been explored for synthesizing qualitative research in software engineering, such as thematic synthesis, meta-ethnography, and meta-summary \cite{h6oJ}. As our aim is to discover patterns or themes of challenges in building products with ML components, as well as get a sense of the priority of the challenges based on the frequency of reports by industry practitioners, the meta-summary method is best suited for this research problem \cite{h6oJ,J0kZ,ZbHh}. The meta-summary method provides a well-balanced synthesis mechanism, which is deeper than mapping studies, and not as exhaustive as meta-ethnography, which requires significant expertise and experience with the methodology and its philosophical stance \cite{J0kZ}. Thus, we apply the meta-summary method \cite{ZbHh,UxiN,J0kZ} to perform the quantitative aggregation of the qualitative evidence that we present as findings. Figure \ref{fig:overview} shows the overview of the research process followed in this study. 

\subsection{Paper Selection }\label{h.dua89v66pr5s}
To increase reliability, reproducibility and objectivity of the process for paper selection, we follow the established procedure of conducting systematic literature reviews \cite{XMGU}. 

\subsubsection{Relevant years}\label{h.efmalr9cxzxk}
Much of the research on engineering products with ML components was inspired by the seminal 2015 ML technical debt paper by Sculley et al. \cite{RlvD}, which outlined various engineering challenges in building and operating ML infrastructure. For completeness, we selected the year range of the papers to be from 2010 to 2022.  

\subsubsection{Publication venues}\label{h.7rrt8vebhz88}
To search for papers, we select digital libraries and databases commonly used by software engineering review papers, e.g, \cite{9igj,pV6m,3bBp}. We do not filter by the venue, as we expect to find papers that are published in different communities including software engineering, human-computer interaction, and machine learning. Since we aim to aggregate results from robust empirical studies, we did not include gray literature, such as blog posts, which typically reflect opinions or individual experience only. However, we did include arXiv as a data source, as it contains many relevant academic papers in this field, even if some have not been peer reviewed. Specifically, we use the following 8 data sources: IEEE Xplore (ieeexplore.ieee.org), ACM Digital library (portal.acm.org/dl.cfm), Wiley InterScience (\url{www.interscience.wiley.com}), Elsevier Science Direct (\url{www.sciencedirect.com}), SpringerLink (\url{www.springerlink.com}), EI Compendex (\url{www.engineeringvillage.com}), and arXiv (\url{https://arxiv.org}).

\subsubsection{Search query}\label{h.qx4jpilg1cww}
Defining the right scope and corresponding search query required some iteration. We started by assembling an initial set of 21 papers as a seed set (a common practice \cite{ffiu,G64S}). The seed set was composed of papers that we knew well from our past work in this field. We then analyzed the seed set to define the keywords needed to retrieve those and similar papers. 

We realized that our research question has three aspects, and therefore to retrieve the papers that would satisfy our research question, we focused on those three parts to formulate the search query: (A) \textbf{The paper needs to mention an ML-related keyword}, since we focus on challenges introduced by ML components. (B) \textbf{The paper needs to mention a software engineering or ML deployment-related keyword}, since we focus on engineering challenges that go beyond local concerns of data scientists; for example the paper should discuss concerns related to actual product development where models are deployed and incorporated into larger software systems. Finally, (C) \textbf{the paper needs to mention surveys or interviews}, since we are interested in the challenges mentioned by industry practitioners and these are the most common relevant research methods; we are not interested in a single-team case study or ethnographic study, as the challenges found in such papers may be specific to individual products.

After adding some semantically similar terminologies, we developed the following search query fragments -- A: ``machine learning'' OR ``artificial intelligence'' OR ``deep learning'' OR ``ML component'' OR ''data science''; B: ''software engineering'' OR ``software systems'' OR ''production-ready systems'' OR ''ML systems'' OR ``deploying ML'' OR ``ML deployment''; C: ``interview'' OR ``survey'' OR ``questionnaire''. The final query was of the following format ``A AND B AND C.''

We searched with this query within the abstract of the papers in all the digital data sources except SpringerLink, as it did not have the option to search within abstracts. For SpringerLink, we retrieved 5612 papers based on a full-text  search, and subsequently used a custom script to search within the abstracts of these papers. This provided us with a total of 341 papers from all the sources (see Table \ref{paperselection}). 

\tableA

This search query retrieved 18 of the 21 seed papers. Two papers were missed because the conducted interviews were not mentioned in the abstract (the abstract framed the research as a case study), and one paper was not listed within the libraries searched (only available on TechRxiv). To account for this difference we performed one round of snowballing, as explained later in this section. 

\subsubsection{Selection criteria}\label{h.7rnuvl792ko0}
The initial search returned many papers that were not directly relevant to our research question. Next, we selected 86 relevant papers by reading the title and abstract, evaluating them against the inclusion and exclusion criteria (see Table \ref{criteria}), which we incrementally refined. Finally, we read the full paper, and once again evaluated each against the inclusion and exclusion criteria, which narrowed our set down to 39 papers. Multiple researchers participated in this process and discussed papers at the boundary. 

Most of the papers that were discarded in this round were either literature surveys in the domain of \emph{machine learning for software engineering} (i.e., using ML techniques to facilitate software engineering tasks; not relevant to this study) or used interviews or surveys to evaluate tools. We also removed papers that have a narrow focus or are entirely model-centric, e.g., interviewing only data scientists about their modeling work (e.g., \cite{ZrtG,b6IF,qOwU,AfqY}) or interviewing only non-technical people (e.g., \cite{3md1,ERdQ,y7Bt,HQ7E}).

\tableB

\subsubsection{Snowballing}\label{h.u5xblheb820c}
To capture relevant papers that did not match our keywords in their abstract, we performed one iteration of backward snowballing \cite{oFbx}, which means that we went through the selected papers' reference list to find whether we missed any relevant papers. We analyzed 26 additional papers and considered 11 of them as relevant based on the inclusion and exclusion criteria, which included the three papers from the seed set we previously missed. 

\subsubsection{Final paper set}\label{h.nkhpa1fvg55f}
Overall, our process resulted in a final set of 50 papers. Most of the papers were published recently, since 2019 (see Figure \ref{fig:yeardist}). This sudden explosion of interview and survey studies with practitioners in recent years justifies our motivation for this study to aggregate all the findings of these papers. Most of the papers, 30 out of 50, were published in software engineering venues (including five at WAIN/CAIN), 11 papers in HCI venues, two papers in AI Ethics venues, and the seven remaining ones are scattered over other communities. A total of 947 interviews and 3811 survey responses were reported in 43 papers, and the seven remaining papers did not report specific counts of the interviewed or surveyed practitioners. 

Of the 50 papers, 31 papers explicitly list research questions or the aim of their research as identification of challenges (or issues, problems, difficulties) in different aspects of building products with ML components. The other papers do not explicitly set a goal of identifying challenges but more broadly study the process of building products with ML components, yet they also report practitioner challenges in their findings. 

\figureB

\subsection{Qualitative Meta-Summary Process}\label{h.q5ux26f3i9vl}
As stated earlier, we used the meta-summary research method \cite{ZbHh,J0kZ,UxiN} to synthesize the findings from the collected papers.  This method is used to perform quantitative aggregation of qualitative findings, which are necessarily the thematic summaries of the underlying data from different studies. We conduct the following steps to perform the synthesis, as per the guidelines.

\subsubsection{Extracting findings}\label{h.q3y7qbengty0}
Along with the standard metadata (title, source, venue, year, etc.), we extracted study-specific data regarding research questions, study method, interview and survey participant counts, and, most importantly, the challenges reported within the papers. To maintain  consistency in extracting the findings, we considered only \emph{challenges} that were derived from the interview and survey answers in the papers, not challenges derived from other literature or personal experience of the authors. We extracted challenges related to building software systems with ML components, but excluded those that relate exclusively to the data- and model-related work performed by a data scientist, such as algorithmic problems, notebook coding, and hyper-parameter tuning. We extracted a total of 520 excerpts relating to challenges from the 50 papers. We stored all extracted information from each paper in a spreadsheet for further analysis. 

\subsubsection{Grouping topically similar findings}\label{h.conawaz7uvxx}
We organize the findings at the level of \emph{reported challenges} that we extracted from the papers. Different papers grouped findings in different ways and using different terminologies; we aimed to find a consistent organizational principle. For identifying similar topics and grouping those together, we needed to understand and compare those reported challenges in their original context. \emph{Card sorting} is a common technique for grouping similar findings \cite{bBfc,9tGL}, which we used for this paper. Following the standard \emph{card sorting} method, we created one (virtual) card per reported challenge, and incrementally and iteratively organized those cards into groups of similar challenges. Multiple researchers went through all the cards in synchronous and asynchronous fashion to grasp the different concepts and identify relevant themes and clusters around the reported challenges. This being a collaborative effort, we did not aim for inter-rater agreement between independent grouping by individual researchers, but instead worked together as a team to build consensus. There were many rounds of card sorting including moving the cards back and forth between different clusters, splitting the cards to handle different dimensions, merging similar clusters, and splitting clusters when we found there was more than one theme, until all involved researchers were satisfied with the clusters and placement of the cards. We developed three layers of clustering -- the reported challenges extracted from the papers as the smallest unit, groups of common \emph{themes} or patterns in the challenges as the second layer, and finally a third (or top) layer grouping the second layer clusters by development stages or cross-cutting concerns for the ease of reporting results. We performed this card sorting process in an online platform (miro.com), allowing us to manipulate colors, add different tags to the cards, add comments, emojis, and so on. We share the resulting card-sorting board as supplementary documents \cite{1JnV}.

\subsubsection{Abstracting and formatting findings}\label{h.812kqcej1mj5}
For each of the second layer clusters we abstracted out the concrete details of the reported challenges and summarized the clusters based on the identified themes of the groups. For this, we once again looked into the cards of each of the clusters individually and attempted to develop broad statements that capture the content of the cards in that cluster, which provide the headings of our results presented in Section \ref{result}. We wanted to be concise, but also comprehensive to properly capture the themes in the card. At the same time, as Sandelowski and Barroso suggested [15], we were careful to preserve the context in which the findings appeared by going back and forth in the original papers when confusion arose, moving cards to other clusters or themes as needed.

\subsubsection{Calculating effect sizes}\label{h.oylmeebjtipb}
Methods for meta-summaries recommend reporting the frequency of findings in the original sources \cite{ZbHh}. Since many of our analyzed papers ask similar broad research questions, we can carefully interpret findings mentioned more frequently as more common, though some papers clearly specialize in specific sub areas such as fairness or software architecture \cite{O0Hy,5GnQ}. We do not attempt to count frequencies of mentions within the papers (``intensity effect size'') because they are not consistently reported, but just report the percentage of papers reporting on a challenge theme (``frequency effect size''). 

\subsection{Limitations and Threats to Validity}\label{h.lehp0uhb3w7p}
All research designs come with limitations that threaten validity and credibility of results. As usual, readers should be careful when generalizing findings beyond what is allowed by the methods. Despite best efforts in our selection methods (SLR process, snowballing) we may have missed some relevant papers. In setting clear rules for scope, we had to do some judgment calls by consensus of all researchers for a number of papers, for example, whether to include \cite{ZrtG,HaFJ,VlSL,UjDH,7fVO}. 

As discussed earlier, the meta-summary synthesis method was chosen deliberately for its fit,but comes with its own limitations: it does not analyze original raw data, but only what is reported by other papers. Organizing and categorizing the data required some interpretation of the papers and some judgment calls. The method encourages quantification of effect sizes, but those may not be entirely reliable as the analyzed papers use different methods and sometimes focus on specific subquestions. 

It would have been interesting to analyze findings in additional dimensions, for example, whether team members in different roles or projects, or in different application domains, experience different challenges, or whether different challenges surface depending on the research method in the original study (e.g., survey vs. interview, open question vs. closed question). Unfortunately, data in the original studies is frequently not reported consistently and with enough granularity to enable such analyses. 

While the meta-summary method can in principle also identify conflicts within the literature, this was not feasible in our study. The analyzed papers typically reported challenges, not the absence or relative importance of certain challenges. Given that different papers often had a different focus, rather than being replications of each other, we cannot conclude that not mentioning a challenge implies that there was no such challenge. Hence, we limited our analysis to aggregating and grouping reported challenges.

\section{Results}\label{h.d4rjik4nls1c}
\label{result}

We report our findings of the meta-summary in this section using the layers derived from the card sorting. The top layer includes development stages (1) \emph{Requirements Engineering}, (2) \emph{Architecture, Design, and Implementation} (with a special focus on (2a) \emph{Model Development} and (2b) \emph{Data Engineering}), and (3) \emph{Quality Assurance,} plus (4) \emph{Process} challenges and (5) \emph{Team} challenges as crosscutting concerns. Also, although \emph{MLOps}, \emph{Fairness}, and other more specific categories are often used to organize results in the surveyed papers, we eventually settled on minimizing the number of cross-cutting topics. We decided to include operations challenges in the \emph{Architecture and Design} group, as we consider them primarily as a \emph{design for change} issue; and we separate and group various concerns for specific qualities, such as \emph{fairness,} in the development stages where the concerns arise, such as requirements and quality assurance. Within these top layer headings, we have our second layer clusters which are the abstracted challenges based on our identified themes, reported as the sub-headings in the following sections.

\subsection{Requirements Engineering}\label{h.kcpfsduxllgw}
Requirements engineering is known as an important and challenging stage of any software project, but as a consistent theme, we find that practitioners argue that the incorporation of ML further complicates requirements engineering.

\subsubsection{\textbf{Lack of AI literacy causes unrealistic expectations from customers, managers, and even other team members} \cite{UjDH,Dk4s,2f5l,DNDN,TtFP,DhfH,kw6g,IrZw,QNQY,xMO3,Yy9q,zo5R,6Iam,XFkw,g6Jy,rk7b,auDR} \textbf{(17/50)}}\label{h.ycbio811ztxp}
Across many studies, many practitioners report that customers frequently have unrealistic expectations of ML capabilities in a product, like demanding a complete lack of false positives or expecting very high accuracy that is infeasible with provided resources (e.g., data, funding). Commonly, practitioners similarly blame a lack of \emph{AI literacy} on customers not wanting to pay for the continuous improvement of the model: they have a static view of model development \cite{QNQY,xMO3} only consider paying for coding, as they do not understand the need for experimental analysis \cite{6Iam} and even difficulty convincing engineering teams to invest in collecting high-quality data \cite{TtFP}. The issue of unrealistic requirements does not only come from customers, but also from team members within the company itself: Data scientists find it hard to explain the capabilities of ML to managers, requirements engineers, and even designers \cite{IrZw,xMO3,zo5R,XFkw,rk7b,auDR}. According to practitioners, a lack of AI literacy in team members manifests particularly in defining and scoping the project: Stakeholders find it hard to understand the suitability of applying ML itself \cite{2f5l,Yy9q}, scoping and deciding the functional and non-functional requirements \cite{DNDN,6Iam}, interpreting the model outcomes \cite{Dk4s,XFkw,g6Jy}, and the infrastructure needs (e.g., appropriate data, monitoring infrastructure, retraining requirements) when building products \cite{DNDN,Yy9q,XFkw}. Many practitioners also report that ML-specific system-level qualities like fairness and explainability are frequently ignored during requirements elicitation, as the stakeholders are not aware of them \cite{DNDN,xMO3,aqp3,SKJO}.

\subsubsection{\textbf{Vagueness in ML problem specifications makes it difficult to map business goals to performance metrics} \cite{MTwb,gqhD,R4n4,Dk4s,2f5l,DNDN,kw6g,IrZw,xMO3,aqp3,O0Hy,6Iam,XpO1,XFkw,g6Jy,x37l,auDR} \textbf{(17/50)}}\label{h.uc13zsbak34m}
Practitioners across many studies mention the challenge of formulating the specific software and ML problem in a way that satisfies business goals and objectives. ML practitioners find it difficult to map the high-level business goals to the low-level requirements for a model. While customers are broadly interested in improving the business, practitioners often find it difficult to quantify the contribution of the ML model and its return on investment. Also, \emph{Responsible AI} initiatives find it difficult to quantify their contributions to the business, for example, measuring the value added by improving fairness and explainability, or to deliberate about tradeoffs between conflicting fairness and business objectives \cite{SKJO,aqp3,O0Hy,auDR}. Even with some notion of the responsible AI requirements in hand, practitioners find the requirements vague and not concrete enough to actually implement (e.g., unclear subpopulations and protected characteristics to balance discrimination) \cite{DNDN,aqp3}. On the other hand, practitioners also frequently report that many projects are exploratory without clear upfront business goals, thus, starting off the project without clear requirements is pretty common, albeit often problematic \cite{Dk4s,IrZw,6Iam,XpO1}. 

\subsubsection{Regulatory constraints specific to data and ML introduce additional requirements that restrict development \cite{XpO1,7fVO,DNDN,SKJO,zo5R,R6PE,Dk4s} (7/50)}\label{h.myyinjotp8nr}
Practitioners in multiple studies expressed how regulatory restrictions constrain ML development and require audits and involvement from legal teams. Privacy laws such as GDPR impose additional requirements on ML practitioners such as ensuring the collection of individual consent \cite{DNDN,zo5R} and providing the nontrivial ability to  remove individuals from training data after they revoke consent. Similarly, practitioners in regulated domains report a need for explainability and transparency that prevents them from using deep learning and post-hoc explainability techniques \cite{7fVO,XpO1,SKJO}.

\subsection{Architecture, Design, and Implementation}\label{h.qp0m47ndeakd}
We find that many ML practitioners struggle with designing the architecture of products with ML components.

\subsubsection{\textbf{Transitioning from a model-centric to a pipeline-driven or system-wide view is considered important for moving into production, but a difficult paradigm shift for many teams} \cite{5GnQ,Dk4s,2f5l,DhfH,Yy9q,6Iam,W82W,x37l,jmTM,aJfF,VItT} \textbf{(11/50)}}\label{h.j9cas9uphdqj}
Practitioners frequently report challenges in migrating from exploratory model code, often in a notebook, to deployable production-quality code in automated ML pipelines \cite{Yy9q,6Iam}. Building an end-to-end ML pipeline is considered to be a challenge due to the difficulties of integrating various ML and non-ML components in a system operating within an environment \cite{5GnQ,Dk4s,2f5l}, the overwhelming complexity of integrating many tools and frameworks \cite{DhfH,x37l,aJfF}, the need for engineering skills beyond the comfort zone of some data scientists \cite{aJfF}, and so on. While practitioners emphasized the importance of pipeline automation for many projects where frequent re-training and deployment of models are needed, they also consider it time-consuming, labor-intensive, error-prone, and not well supported by current tools \cite{Dk4s,W82W,x37l,jmTM,VItT}.

\subsubsection{\textbf{ML adds substantial design complexity with many, often implicit, data and tooling dependencies, and entanglements due to a lack of modularity} \cite{Dk4s,4myf,x37l,MTwb,gqhD,5GnQ,IrZw,Yy9q,jmTM,KOCv,rk7b} \textbf{(11/50)}}\label{h.uks6ns9g1a4f}
Many practitioners report challenges from additional complexity when designing systems incorporating machine learning, and the traditional software architecture and design practices no longer fit  \cite{gqhD,5GnQ,Yy9q,rk7b}. ML changes the assumptions in traditional software systems such as encapsulation and modularity and causes entanglements of data, source code, and ML models, which can lead to \emph{``pipeline jungles''} and \emph{``change anything changes everything''} integrations that are hard to maintain  \cite{Dk4s,x37l,gqhD,IrZw,4myf,jmTM}. Unlike traditional systems, ML requires the incorporation of data pipelines that need to handle a high volume of data and often data architectures of distributed nature, and practitioners also need to understand and design for the data flow in the entire system \cite{MTwb,IrZw,Yy9q}. Practitioners also point out that complexities arise due to a large amount of surrounding ``glue code'' to support the ML models \cite{Dk4s,KOCv}, and complicated dependency and configuration management \cite{IrZw,KOCv}. 

\subsubsection{\textbf{Difficulty in scaling model training and deployment on diverse hardware} \cite{Dk4s,DhfH,6Iam,wHQl,XpO1,g6Jy,4myf,x37l,VItT,aJfF} \textbf{(10/50)}}\label{h.n2z37oroj03n}
Practitioners commonly report difficulty dealing with cloud and computational resources, even with the recent emergence of MLOps. Practitioners find the technologies to be difficult to integrate into the production environment and require substantial time, effort, and money \cite{DhfH,6Iam,XpO1,g6Jy,x37l,aJfF}. Among the common problems of such deployments, practitioners brought up the mismatch of development and production environments \cite{6Iam,4myf}, difficulties in building a scalable pipeline \cite{wHQl,XpO1,x37l,VItT}, adhering to serving requirements such as latency and throughput \cite{Dk4s,x37l}, as well as undocumented tribal knowledge within the team, hampering future deployments \cite{4myf}. Despite the emerging MLOps tooling, practitioners still raise many questions about how to utilize those resources and sometimes express being overwhelmed by the sudden flood of tools and frameworks to choose from \cite{TtFP,HaFJ}.

\subsubsection{\textbf{While monitorability and planning for change are often considered important, they are mostly considered only late after launching} \cite{Dk4s,XpO1,4myf,VItT,5GnQ,R4n4,Yy9q,jmTM,2f5l,omzN,xMO3,KOCv,SKJO,fCMZ,aJfF} \textbf{(15/50)}}\label{h.7gdn3bizsxqy}
Practitioners report struggling with monitoring their deployed models for detecting drift, bias, or even failures. While many highlight monitoring as very important, planning for monitoring is rare \cite{xMO3}. Even for companies that adopt a monitoring infrastructure, practitioners report struggling with ad-hoc monitoring practices of logging, creating alerts, or doing everything manually \cite{4myf,5GnQ}. Similar concerns were raised about model evolution, where practitioners acknowledge it to be important, but fall behind in planning for change in their architectural design \cite{Dk4s,XpO1,VItT,jmTM,KOCv}. Practitioners mentioned that ML-centric software goes through frequent revisions more than traditional software  (e.g., due to model retraining, or even model replacement for data change, hyperparameter tuning, or change of domain, etc.), and the changes tend to be nontrivial and nonlocal, raising the need for an architecture that supports such changes. As a result, we find practitioners' soliciting the need for adapted architectural patterns to design for such post-launch activities for products with ML components with monitorability as a significant quality attribute \cite{5GnQ,Yy9q}. 

\subsection{Model Development}\label{h.qr27oaroeq01}
Although we explicitly exclude challenges relating only to the work and tools of data scientists when building models, we find reports of engineering challenges during model development, which we report in this section.

\subsubsection{\textbf{Model development benefits from engineering infrastructure and tooling but provided infrastructure and technical support are limited in many teams} \cite{6Iam,XpO1,VItT,R4n4,IrZw,2f5l,omzN,KOCv,kw6g,TtFP,R6PE,IjZD,ke7w,CN5W,W82W,GZrM,XFkw,HaFJ,VlSL} \textbf{(19/50)}}\label{h.bb9u2eegvcln}
ML practitioners share tooling needs for different tasks including data analysis and visualization, feature engineering, model development, integration, evaluation, deployment, monitoring, reproducibility, and support for specific qualities like privacy, security, and explainability. They report a lack of adequate tools in these areas and find the existing tools and techniques to be (a) unavailable in their environment \cite{XpO1,IjZD}, (b) not automated enough \cite{ke7w}, (c) requiring too much expert knowledge to be used \cite{CN5W,R4n4,R6PE,ke7w}, (d) limited to specific tasks and types of data sets \cite{IjZD,ke7w}, or (e) not suitable for their own problems \cite{IjZD,VlSL,TtFP}. This raises demand for custom tools but many teams lack the resources and engineering support.

\subsubsection{\textbf{Code quality is not standardized in model development tools, leading to conflicts about code quality} \cite{4myf,IrZw,xMO3} \textbf{(3/50)}}\label{h.eu4vggho428f}
Practitioners report that code quality and review processes are usually not standardized and are inconsistent across development and production environments. The expectations around code quality and versioning also differ widely in teams and create conflicts within teams, especially among team members with different roles and backgrounds. Practitioners commonly complain about low code quality in data science code, especially in notebooks.

\subsection{Data Engineering}\label{h.fu7zmix73ao4}
In developing machine learning models, data plays an important role. While we exclude challenges related exclusively to data-related work within ML pipelines, we report engineering challenges related to handling data within the system.

\subsubsection{\textbf{Data quality is considered important, but difficult for practitioners and not well supported by tools} \cite{Dk4s,DhfH,6Iam,XpO1,g6Jy,4myf,x37l,gqhD,jmTM,xMO3,KOCv,DNDN,TtFP,zo5R,ke7w,GZrM,aJfF} \textbf{(17/50)}}\label{h.b0xrlctn33j7}
ML practitioners commonly report struggling with validating and improving data quality. Even with significant research efforts in building tools for data labeling, cleaning, visualization, and management, data work is still reported as a problematic area for practitioners. Practitioners reported that they need to invest significant effort and time in data pre-processing, cleaning, and assembly \cite{GZrM,zo5R,xMO3,jmTM,ke7w,x37l,6Iam,TtFP}. Practitioners also mention their pain points in handling data errors and validating data quality, where better tool support is desired \cite{wHQl,Dk4s,4myf,TtFP,g6Jy,DNDN,gqhD,aJfF}. Although it is common to associate these data issues within the model building pipeline, practitioners feel the need for cooperation from other parts of the organization (e.g., requirements engineers need to identify and specify requirements regarding data collection, formats, and the ranges of data and domain experts need to help to understand the structure and semantics of the data), which they mention is lacking \cite{g6Jy,DNDN,ke7w,xMO3,zo5R}.

\subsubsection{\textbf{Internal data security and privacy policies restrict data access and use} \cite{DhfH,6Iam,XpO1,g6Jy,x37l,2f5l,xMO3,KOCv,TtFP,qky1} \textbf{(10/50)}}\label{h.js04j7cgfolo}
Data access is often restricted due to security and privacy policies within organizations, beyond possible regulatory restrictions, e.g., policies ensuring that customer data is not shared outside the company. Due to restrictions on the flow of data, ML practitioners need to deal with additional complexities in the data pipeline, as only a restricted number of team members can analyze the data and as they have limited access to the right data and no access to data locally for model optimization or model debugging due to data movement constraints \cite{qky1,6Iam,x37l,XpO1}. 

\subsubsection{\textbf{Although training-serving skew is common, many teams lack support for its required detection and monitoring} \cite{XpO1,4myf,x37l,R4n4,Yy9q,xMO3,KOCv} \textbf{(7/50)}}\label{h.85g3touusqzn}
The mismatch between training data and production data is a common problem in products with ML components, where models work well on test data but generalize poorly to real-world data in production. Even if training the model with a representative dataset initially, the production environment often encounters drift toward data distributions that are less well supported by the model. Practitioners explain that monitoring models in production for staleness is an important activity that supports detecting the degradation of model performance and retraining it with new data if needed. However, they also find it challenging to set up the monitoring infrastructure and report a lack of tool support.

\subsubsection{\textbf{Data versioning and provenance tracking are often seen as elusive, with not enough tool support} \cite{DhfH,wHQl,VItT,jmTM,2f5l,kw6g,zo5R} \textbf{(7/50)}}\label{h.my1eetaef92r}
While software engineers routinely adopt  mature version control systems for code, practitioners report challenges in versioning data, typically due to the large volumes of data involved. Practitioners mention that they need to have traceability and transparency to answer questions like  \emph{``Which data was this model trained on?''} or \emph{``Which code or data change made our accuracy deteriorate?'' \cite{VItT}}, but it's not possible for them to keep track of data and models across the life cycle without technological support  \cite{VItT,jmTM,DKy1}. This is a bigger problem for practitioners in small companies as they do not want to invest in storage capacity to version their models and datasets, though they understand the importance \cite{zo5R}.

\subsection{Quality Assurance}\label{h.nxhfittmqk3n}
One of the biggest changes that the incorporation of ML models has brought into traditional software development is challenging the traditional notion of correctness, where models are evaluated for accuracy or fit rather than whether they fully meet a specification. Understandably this impacts the conventional processes and practices associated with testing and quality assurance. 

\subsubsection{\textbf{Testing and debugging ML models is difficult due to lack of specifications} \cite{Dk4s,kw6g,QNQY,ke7w,CN5W,W82W,GZrM,gqhD,R4n4,IrZw,xMO3,6Iam,wHQl,XpO1,g6Jy,4myf,x37l,jmTM,KOCv} \textbf{(19/50)}}\label{h.bqibkzumjghn}
Practitioners find testing and debugging of ML models challenging. In particular, they ubiquitously report difficulty establishing quality assurance criteria and metrics, given that no model is expected to be always correct, but it is difficult to define what amount and what kind of mistakes are acceptable for a model \cite{6Iam,XpO1,x37l,gqhD,IrZw,xMO3,KOCv,QNQY,CN5W,GZrM}. In particular, practitioners find it difficult to define accuracy thresholds for evaluations.  Furthermore, practitioners report finding it difficult to select adequate test data, specifically curating test data of sufficient quality and quantity  that is representative of the production environment \cite{R4n4,IrZw,kw6g,ke7w,W82W}. Curating test data for ML testing is also considered costly and labor-intensive, and practitioners desire methods and tools from the research community for automated test input generation to reduce this cost \cite{gqhD,IrZw,QNQY}. Practitioners consider it a challenge to get labels for test data and evaluate test quality (e.g., in terms of coverage) due to the difficulty of defining the valid input space and the test oracle problem \cite{gqhD,IrZw,GZrM}. Practitioners also mention the silent failing of models (i.e., models give wrong answers rather than crashing), the long tail of corner cases, and the ``invisible errors'', that are handled on an ad-hoc basis without a systematic framework or a standard approach \cite{4myf,IrZw,GZrM}. Additionally, practitioners raise challenges regarding evaluating model robustness, on one hand, suffering from the lack of a concrete methodology \cite{ke7w,GZrM}, and on the other hand, having various metrics but no consensus on which metric to use \cite{gqhD}.

\subsubsection{\textbf{Testing of model interactions, pipelines, and the entire system is considered challenging and often neglected} \cite{x37l,gqhD,2f5l,xMO3,ke7w,CN5W,W82W,GZrM} \textbf{(8/50)}}\label{h.vdkihxvhkneb}
Testing literature often focuses on ML models and data quality, but less on how models are integrated into the system, and even less on the infrastructure to produce the models. Practitioners find sole unit testing of individual models insufficient and ineffective, due to the entanglement of models and different ML components, as well as the difficulty of explaining why an error occurred due to the low interpretability of individual models \cite{gqhD,IrZw,CN5W}. The lack of pipeline and system testing beyond the model is also considered a problematic area \cite{2f5l,xMO3,ke7w,CN5W,W82W,GZrM}: While practitioners tend to focus more on the data- and model-related issues, the error handling around the model is found to be insufficient in previous studies \cite{CN5W,GZrM}, leading to system failures even where the model gives the correct results \cite{W82W}. Practitioners also report having no systematic evaluation strategy nor automated tools and techniques for pipeline and system-level testing \cite{gqhD,xMO3}.

\subsubsection{\textbf{Testing and monitoring models in production are considered important but difficult, and often not done} \cite{Dk4s,4myf,gqhD,xMO3,ke7w} \textbf{(5/50)}}\label{h.16ag52an0arp}
Many practitioners recognize the need to test in production (online testing), since offline test data for models may not be representative, especially as data distributions drift. However,  practitioners consider online testing complex as it is not trivial for them to design online metrics that do not only depend on the model but also on the external environment, user interactions after deployment, and the context of the product overall \cite{4myf,gqhD}. Practitioners also find online testing very time-consuming, as it requires longer observation periods to determine meaningful results \cite{Dk4s,4myf}. Practitioners also pointed out that there is no surefire strategy to precisely detect when the model is underperforming in online testing \cite{4myf}.

\subsubsection{\textbf{There are no standard processes or guidelines on how to assess system qualities such as  fairness, security, and safety in practice} \cite{VItT,omzN,O0Hy,kw6g,zo5R,fCMZ,7fVO,SKJO,VlSL} \textbf{(9/50)}}\label{h.5wwnlupjafi7}
Research often discusses how machine learning influences fairness, robustness, security, safety, and other qualities, but practitioners report that they find evaluating these as challenging. While practitioners consider these qualities important \cite{VItT,omzN}, they often report having no effective methodology or concrete guidelines for evaluating them \cite{omzN,O0Hy,kw6g,zo5R,fCMZ,7fVO,VlSL}. Even regarding fairness, which has received a lot of research attention lately, practitioners report finding it hard to apply auditing and de-biasing methods due to not having a proper process in place \cite{O0Hy,zo5R}. Some practitioners report waiting for complaints from customers rather than being proactive when it comes to fairness \cite{O0Hy}, or even blindly expecting the algorithms to inherently provide qualities like security against attacks \cite{omzN}.

\subsection{Process}\label{h.kha9tgompi5l}
Building software products with ML components involves many moving parts that need to be planned and integrated. Fitting all of these together in a cohesive process can be challenging. 

\subsubsection{\textbf{Development of products with ML component(s) is often  ad-hoc, lacking well-defined processes}  \cite{6Iam,XpO1,MTwb,IrZw,2f5l,KOCv,kw6g,QNQY,TtFP,W82W,UjDH} \textbf{(11/50)}}\label{h.8nnropdye2tx}
Many practitioners report that they struggle with finding a good process for developing ML components and products around them \cite{6Iam,XpO1,MTwb,IrZw,QNQY}, often coming up with ad-hoc strategies and experiencing a lack of good engineering practices \cite{QNQY,W82W}. ML practitioners have explored using the traditional software development life cycles and found those to be a poor fit for exploratory development work. Even with a flexible agile methodology, practitioners identified that small iterations of sprints cannot fit the initial feasibility study that ML requires, with the timeline being too fixed and too short \cite{6Iam,XpO1,UjDH}. Also, they find it hard to set expectations for each sprint, as the project objectives may remain unclear at the beginning and need to be revisited after the initial investigation \cite{2f5l,UjDH}.

\subsubsection{\textbf{The uncertainty in ML development makes it hard to plan and estimate effort and time} \cite{6Iam,IrZw,KOCv,kw6g,QNQY,GZrM,UjDH} \textbf{(7/50)}}\label{h.rzpl0r295ekg}
Machine learning work tends to be iterative and exploratory and as such uncertain, where practitioners cannot estimate upfront how long it may take to reach a model with a certain level of accuracy or whether that is even possible at all; instead, they commonly progress with many experiments with different algorithms and datasets \cite{KOCv,QNQY}. Practitioners, therefore, report having difficulties setting expectations and (intermediate) deadlines for a project \cite{6Iam,IrZw,KOCv,GZrM,UjDH} and providing any upfront estimates about effort and cost \cite{6Iam,KOCv,kw6g}.

\subsubsection{\textbf{Practitioners find documentation more important than ever in ML, but find it more challenging than traditional software documentation} \cite{6Iam,wHQl,XpO1,R4n4,xMO3,cupz,SCCw,DKy1,QOKm} \textbf{(9/50)}}\label{h.slari7qkxifk}
Many practitioners point out various process and coordination challenges rooted in poor documentation. Some practitioners emphasize that documentation is even more important when it comes to ML components, as human decisions are inscribed in different stages of ML pipelines and cannot be retrieved from code or data without documentation \cite{XpO1,DKy1}. The final model code is the outcome of many different explorations and experimentations that include multiple rounds of data processing, feature engineering, hyperparameter tuning, and other activities. Many problem-specific decisions have been made in those stages that cannot be understood from the resulting model or pipeline code. Some argue that not recording these decisions in documentation causes them to slowly become invisible, severely impacting future re-analysis and revisions, or even model integration and deployment \cite{XpO1,R4n4,DKy1}. Others emphasize that, along with model documentation, data documentation is also imperative to share hidden information inside the data and create a shared data understanding, yet mostly missing in organizations \cite{wHQl,xMO3}. Others report that, with the incorporation of ML, the documentation process becomes more complicated as ML practitioners find it difficult to present complex model information in an accessible way to all levels of stakeholders \cite{cupz,SCCw,QOKm}. It is also non-trivial for practitioners to decide on the right amount of details to include in the documentation. They place the blame mostly on the lack of organizational incentives, resources, and unclear and vague guidelines for ML documentation \cite{cupz,QOKm}.

\subsection{Organization and Teams}\label{h.ai9t7gj6mrpk}
Along with the challenges faced in different development stages, practitioners also mention challenges they suffer from the organizational and teamwork perspective while building products with ML components. 

\subsubsection{\textbf{Building products with ML components requires diverse skill sets, which is often missing in development teams} \cite{Dk4s,DhfH,IrZw,Yy9q,KOCv,kw6g,AaNn,qky1,W82W,XFkw,UjDH,HaFJ} \textbf{(12/50)}}\label{h.9iwyszxn0yw5}
Incorporation of ML in a product does not merely mean adding just another component to the system; it requires people from multiple disciplines to get involved to support different aspects of this component. The team requires many diverse skill sets to develop, deploy, and integrate the model into the complete product, including hardware expertise, engineering skills, knowledge of math and statistics, business understanding, UX design ability, operations, and domain expertise.  The lack of this varied expertise in the team is commonly mentioned to be a challenge by practitioners \cite{Dk4s,IrZw,kw6g,qky1,UjDH,HaFJ}. Also, as discussed in the next subsection that communication is often hindered by a lack of AI literacy or common terminology \cite{XpO1,jmTM,KOCv,W82W,XFkw,Yy9q}, cross-disciplinary knowledge seems to be important for team members to interact and understand each other's vocabulary; however, practitioner experiences seem to indicate that such cross-disciplinary education is not broadly available yet \cite{4myf,GZrM}.

\subsubsection{\textbf{Many teams are not well prepared for the extensive interdisciplinary collaboration and communication needed in ML products} \cite{DhfH,wHQl,Yy9q,xMO3,KOCv,IjZD,W82W,XFkw,a19O,rk7b,IrZw} \textbf{(11/50)}}\label{h.17h2u1ngpcym}
For building a product with ML components, team members need to collaborate with people from different disciplines as mentioned above, such as business leaders, engineers, designers, and various other departments inside the company, and even outside the organization \cite{wHQl,IrZw,Yy9q,XFkw,rk7b}.  Practitioners report that they often struggle to collaborate effectively in such interdisciplinary teams, because team members often do not understand the concerns of other members from other backgrounds, like data scientists lacking knowledge of engineering practices, testing frameworks, continuous integration and delivery, and such \cite{XpO1,jmTM,KOCv,W82W}; software engineers lacking AI literacy \cite{XpO1,W82W}; and data scientists and software engineers not understanding or interacting members with in with business roles \cite{Yy9q,XFkw}. Practitioners report struggling with cultural differences, differences in expectations, and conflicting priorities \cite{DhfH,KOCv,IjZD,XFkw}, and they often do not agree on assigned responsibilities \cite{6Iam,xMO3}. These multidisciplinary teams also suffer from miscommunications arising from inconsistency in their technical terminologies \cite{xMO3,W82W,a19O}. Siloing of teams by specialization and lack of communication across such silos are also observed in many production settings, fostering integration problems even further \cite{xMO3,IjZD}. 

\subsubsection{\textbf{ML development can be costly and resource limits can substantially curb/limit efforts} \cite{DhfH,4myf,KOCv,qky1,XFkw,HaFJ} \textbf{(6/50)}}\label{h.wsx0m8rsmeyh}
Practitioners report that organizations involved in the development of products with ML components often suffer from resource and budget limitations. Hardware, infrastructure, cloud storage, GPUs, etc., are expensive, and especially for small companies, it is difficult to justify such expenditures based on the expected return on investment from the model.

\subsubsection{\textbf{Lack of organizational incentives, resources, and education hampers achieving all system-level qualities} \cite{omzN,aqp3,R6PE,qky1,fCMZ,XFkw,7fVO,VlSL} \textbf{(8/50)}}\label{h.ebog29ttkqkv}
Practitioners mention that organizational incentives also have an impact on achieving certain qualities of products with ML components. A quality that practitioners reported frequently as particularly challenging due to the lack of organizational incentives is fairness \cite{aqp3,O0Hy}. Awareness of potential problems, including potential consequences from biased models, seems to be the main reason for lacking responsible AI practices, along with the lack of organizational incentives and structures, as well as priority conflicts. Safety, security, and privacy also seem to suffer from similar issues of awareness, education, resource constraints, and are often disregarded due to tradeoffs with development cost \cite{omzN,R6PE,qky1,fCMZ,VlSL}.

\section{Discussion and Conclusions}\label{h.hz5wmidlusqz}
With this meta-summary, we aggregate and summarize the challenges reported by industry practitioners who build software products with ML components. We find that practitioners report challenges in all stages of the development process, from the initial requirements specification stage to quality assurance of the deployed product. They report a broad range of issues from lacking process, organizational structure, and team collaboration strategies, to lacking tool support for data, model building, deployment, and monitoring. 

\subsubsection{Old, new, and harder challenges}\label{h.p5v5k0gngzvb}
Arguably, many reported challenges are not new to software engineers, and likely many software engineers may have reported similar challenges in non-ML projects. It seems though that the introduction of machine learning exacerbates some universal challenges and introduces new ones. For example, software engineering literature is well aware that requirements engineering is challenging, with customers having unrealistic expectations and developers directly jumping into coding without understanding requirements first. While our study does not support direct comparisons, it seems that these problems haunt ML practitioners more, given how ML inspires hopes for amazing capabilities, but in a way that may be difficult to understand and specify without substantial ML expertise. Similarly, the software-engineering literature is full of nuanced discussions of development life cycles and competing process models, but ML practitioners struggle adopting even the most flexible agile-inspired processes for their projects with the uncertainty that ML brings. Also, team collaboration and organizational challenges are well known in traditional software engineering, but those seem to become even more central with the additional complexity and inclusion of more people with different backgrounds, cultures, and priorities. Other challenges seem new, such as the data- and model-related challenges associated with ML components, and several of the reported challenges regarding architecture and quality assurance stemming from the different nature of reasoning in machine learning.

\subsubsection{Toward better engineering of ML products}\label{h.6x0ch5pkfisr}
A finding from our study is that there is much more consensus on what the challenges are, than how to overcome them. Some challenges could be addressed with new tooling or new practices; for others it may be possible to simply adopt existing good engineering practices; and yet others may just be intrinsically hard problems. While we cannot provide a rigorous summary or analysis, we close by reflecting on possible directions.

\begin{compactitem}
	\item \textbf{Requirements Engineering.} For the challenges of unrealistic requirements, several studies mentioned that practitioners found it useful to conduct training sessions with clients and other team members on AI literacy, before starting the ML projects \cite{xMO3,J0kZ,FBN2,DNDN}. But again, while many practitioners mention suffering from unclear model requirements, we still do not seem to have a good solution to that, and additional research on how to elicit and describe requirements for models may be needed. Another area for future research would be to better understand and prepare for regulatory constraints and provide evidence of compliance.

	\item \textbf{Architecture, Design, and Implementation.} Machine learning seems to provide significant challenges to architectural design of software systems, but arguably many challenges are similar to other large and complex and distributed software systems. While there are nascent discussions on organizing architecture knowledge as patterns \cite{Dk4s,YY1S,5GnQ,U3jM}, it does not seem like the field has reached saturation. This seems to be a field though, where industry-oriented research (similar to the data architecture of facebook \cite{cHAy}) has more access to the complicated real-world scenarios where architectural planning becomes important than what academics can typically access. From the challenges raised by practitioners, it is apparent that along with the need for design practices, patterns, and mechanisms to handle system and model-level considerations (e.g., dependency management, scalability, monitorability), we also need to support teams in shifting from model-centric work to system thinking, possibly through tailored education for ML practitioners.

	\item \textbf{Model Development and Data Engineering.} Consistent across many papers, we find that ML practitioners desire more engineering support, such as better infrastructure and tools for model and data work. Data scientists also indicate a need for more cooperation from other team members in terms of support for data, which necessitates better collaboration strategies and data education for the entire team. On the other hand, a few practitioners highlighted the necessity of standardization of ML code quality, which may be a low hanging fruit technically, but may require a change to the culture and practices in many projects.

	\item \textbf{Quality Assurance.} Quality assurance for machine learning, especially for models, is a very active area of research, with proposals for many different testing strategies to validate different model characteristics covered in multiple literature surveys \cite{PSLS,InOi,wvQU,hW2b}.  While we found that a lot of practitioners mentioned concerns about specifying model adequacy goals, few practitioners showed concerns about system testing, monitoring in production, and testing for fairness, security, and safety. We are surprised to not see more concerns about system-level quality beyond the model, which might indicate either that practitioners do not consider these testing areas as challenging, or that most organizations (especially outside of big tech) are not yet mature enough to even start thinking about such testing needs. Monitoring though is recognized as an important challenge, with many available tools but common adoption problems that may be worth investigating further.

	\item \textbf{Process.} While there is research on the development processes for ML models \cite{Hi9f,E5GQ}, there seems to be little work on addressing process challenges that arise when integrating ML and non-ML work in production projects that are commonly mentioned by practitioners. We believe that this is an area with plenty of research opportunities to evaluate what processes and practices work well in different contexts.

	\item \textbf{Organization and Teams.} While there is lots of research on technical issues, practitioners often see organizational and team issues (such as a lack of AI literacy in teams, unclear responsibility boundaries, and a lack of team synchronization) as some of the most difficult challenges to overcome. Education and better collaboration strategies seem to be the factors that might put a positive impact on mitigating many of the challenges that the practitioners mentioned.

\end{compactitem}

Overall we believe that a lot of progress can be made with better education and better adoption of good software engineering practices. There are plenty research opportunities to adapt existing practices, support them with tooling, and create new interventions altogether. We hope that the collection of challenges, which can be traced to the original studies where they were raised by practitioners, will be helpful in selecting and prioritizing research and education in our community.

\textbf{Acknowledgments.} Kästner's, Nahar's, and Zhang's work was supported in part by the National Science Foundation (\#2131477), Zhou's work was supported in part by the Natural Sciences and Engineering Research Council of Canada (NSERC, RGPIN2021-03538), and Lewis' work was funded and supported by the Department of Defense under Contract No. FA8702-15-D-0002 with Carnegie Mellon University for the operation of the Software Engineering Institute, a federally funded research and development center (DM23-0228).

\balance

\clearpage
\section*{Appendix}
\subsection*{List of Papers}
\begin{enumerate}[leftmargin=*]
    \item d. S. Nascimento, E., Ahmed, I., Oliveira, E., Palheta, M.P., Steinmacher, I. and Conte, T. 2019. Understanding Development Process of Machine Learning Systems: Challenges and Solutions. \emph{Proceedings of the 2019 ACM/IEEE International Symposium on Empirical Software Engineering and Measurement (ESEM)}, 1–6.
    \item Lewis, G.A., Bellomo, S. and Ozkaya, I. 2021. Characterizing and Detecting Mismatch in Machine-Learning-Enabled Systems. \emph{Proceedings of the IEEE/ACM 1st Workshop on AI Engineering-Software Engineering for AI (WAIN)}, 133–140.
    \item Lewis, G.A., Ozkaya, I. and Xu, X. 2021. Software Architecture Challenges for ML Systems. \emph{Proceedings of the 2021 IEEE International Conference on Software Maintenance and Evolution (ICSME)}, 634–638.
    \item Serban, A. and Visser, J. 2022. Adapting Software Architectures to Machine Learning Challenges. \emph{Proceedings of the 2022 IEEE International Conference on Software Analysis, Evolution and Reengineering (SANER)}, 152–163.
    \item Vogelsang, A. and Borg, M. 2019. Requirements Engineering for Machine Learning: Perspectives from Data Scientists. \emph{Proceedings of the 27th International Requirements Engineering Conference Workshops (REW)}, 245–251.
    \item Kim, M., Zimmermann, T., DeLine, R. and Begel, A. 2018. Data Scientists in Software Teams: State of the Art and Challenges. \emph{IEEE Transactions on Software Engineering}. 44, 11, 1024–1038.
    \item Ishikawa, F. and Yoshioka, N. 2019. How do engineers perceive difficulties in engineering of machine-learning systems? - questionnaire survey. \emph{Proceedings of the 2019 IEEE/ACM Joint 7th International Workshop on Conducting Empirical Studies in Industry (CESI) and 6th International Workshop on Software Engineering Research and Industrial Practice (SER\&IP)}, 2–9.
    \item Nahar, N., Zhou, S., Lewis, G. and Kästner, C. 2022. Collaboration Challenges in Building ML-Enabled Systems: Communication, Documentation, Engineering, and Process. \emph{Proceedings of the 44th International Conference on Software Engineering}, 413–425.
    \item Rakova, B., Yang, J., Cramer, H. and Chowdhury, R. 2020. Where Responsible AI meets Reality: Practitioner Perspectives on Enablers for shifting Organizational Practices. \emph{Proceedings of the ACM on Human-Computer Interaction}, 1–23.
    \item Holstein, K., Wortman Vaughan, J., Daumé, H., Dudik, M. and Wallach, H. 2019. Improving Fairness in Machine Learning Systems: What Do Industry Practitioners Need? \emph{Proceedings of the 2019 CHI Conference on Human Factors in Computing Systems}, 1–16.
    \item Hopkins, A. and Booth, S. 2021. Machine Learning Practices Outside Big Tech: How Resource Constraints Challenge Responsible Development. \emph{Proceedings of the 2021 AAAI/ACM Conference on AI, Ethics, and Society}, 134–145.
    \item Serban, A., van der Blom, K., Hoos, H. and Visser, J. 2020. Adoption and Effects of Software Engineering Best Practices in Machine Learning. \emph{Proceedings of the 14th ACM/IEEE International Symposium on Empirical Software Engineering and Measurement (ESEM)}, 1–12.
    \item Rahman, M.S., Khomh, F., Hamidi, A., Cheng, J., Antoniol, G. and Washizaki, H. 2021. Machine Learning Application Development: Practitioners' Insights. \emph{arXiv [cs.SE]}.
    \item Haakman, M., Cruz, L., Huijgens, H. and van Deursen, A. 2021. AI Lifecycle Models Need To Be Revised. An Exploratory Study in Fintech. \emph{Empirical Software Engineering}. 26, 5, 1–29.
    \item Wan, Z., Xia, X., Lo, D. and Murphy, G.C. 2019. How does Machine Learning Change Software Development Practices? \emph{IEEE Transactions on Software Engineering}. 47, 9, 1857–1871.
    \item Washizaki, H., Takeuchi, H., Khomh, F., Natori, N., Doi, T. and Okuda, S. 2020. Practitioners' insights on machine-learning software engineering design patterns: a preliminary study. \emph{Proceedings of the 2020 IEEE International Conference on Software Maintenance and Evolution (ICSME)}, 797–799.
    \item Li, S., Guo, J., Lou, J.-G., Fan, M., Liu, T. and Zhang, D. 2022. Testing machine learning systems in industry: an empirical study. \emph{Proceedings of the 44th International Conference on Software Engineering: Software Engineering in Practice}, 263–272.
    \item Nikhil, K., Anandayuvaraj, D., Detti, A., Lee Bland, F., Rahaman, S. and Davis, J.C. 2022. ``If security is required'': Engineering and Security Practices for Machine Learning-based IoT Devices. \emph{Proceedings of the 4th International Workshop on Software Engineering Research and Practices for the IoT (SERP4IoT)}, 1–8.
    \item Chang, J. and Custis, C. 2022. Understanding Implementation Challenges in Machine Learning Documentation. \emph{Equity and Access in Algorithms, Mechanisms, and Optimization}, 1–8.
    \item Bäuerle, A., Cabrera, Á.A., Hohman, F., Maher, M., Koski, D., Suau, X., Barik, T. and Moritz, D. 2022. Symphony: Composing Interactive Interfaces for Machine Learning. \emph{Proceedings of the 2022 CHI Conference on Human Factors in Computing Systems}, 1–14.
    \item Laato, S., Birkstedt, T., Mäantymäki, M., Minkkinen, M. and Mikkonen, T. 2022. AI governance in the system development life cycle: insights on responsible machine learning engineering. \emph{Proceedings of the 1st International Conference on AI Engineering: Software Engineering for AI}, 113–123.
    \item Mäkinen, S., Skogström, H., Laaksonen, E. and Mikkonen, T. 2021. Who Needs MLOps: What Data Scientists Seek to Accomplish and How Can MLOps Help? \emph{Proceedings of the IEEE/ACM 1st Workshop on AI Engineering - Software Engineering for AI (WAIN)}, 109–112.
    \item John, M.M., Olsson, H.H. and Bosch, J. 2020. AI Deployment Architecture: Multi-Case Study for Key Factor Identification. \emph{Proceedings of the 27th Asia-Pacific Software Engineering Conference (APSEC)}, 395–404.
    \item Uchihira, N. 2022. Project FMEA for Recognizing Difficulties in Machine Learning Application System Development. \emph{Proceedings of the 2022 Portland International Conference on Management of Engineering and Technology (PICMET)}, 1–8.
    \item Kumar, R.S.S., Nystrom, M., Lambert, J., Marshall, A., Goertzel, M., Comissoneru, A., Swann, M. and Xia, S. 2020. Adversarial Machine Learning - Industry Perspectives. \emph{Proceedings of the 2020 IEEE Security and Privacy Workshops (SPW).}, 69–75.
    \item Zdanowska, S. and Taylor, A.S. 2022. A study of UX practitioners roles in designing real-world, enterprise ML systems. \emph{Proceedings of the 2022 CHI Conference on Human Factors in Computing Systems}, 1–15.
    \item Liu, H., Eksmo, S., Risberg, J. and Hebig, R. 2020. Emerging and Changing Tasks in the Development Process for Machine Learning Systems. \emph{Proceedings of the International Conference on Software and System Processes}, 125–134.
    \item Zhang, X., Yang, Y., Feng, Y. and Chen, Z. 2019. Software Engineering Practice in the Development of Deep Learning Applications. \emph{arXiv [cs.SE]}.
    \item Rismani, S., Shelby, R., Smart, A., Jatho, E., Kroll, J., Moon, A. and Rostamzadeh, N. 2022. From plane crashes to algorithmic harm: applicability of safety engineering frameworks for responsible ML. \emph{arXiv [cs.HC]}.
    \item Myllyaho, L., Raatikainen, M., Männistö, T., Nurminen, J.K. and Mikkonen, T. 2022. On misbehaviour and fault tolerance in machine learning systems. \emph{Journal of Systems and Software}. 183,, 111096.
    \item Golendukhina, V., Lenarduzzi, V. and Felderer, M. 2022. What is software quality for AI engineers? Towards a thinning of the fog. \emph{Proceedings of the 1st International Conference on AI Engineering: Software Engineering for AI}, 1–9.
    \item Königstorfer, F. and Thalmann, S. 2022. AI Documentation: A path to accountability. \emph{Journal of Responsible Technology}. 11,, 100043.
    \item Riungu-Kalliosaari, L., Kauppinen, M. and Männistö, T. 2017. What Can Be Learnt from Experienced Data Scientists? A Case Study. \emph{Product-Focused Software Process Improvement}, 55–70.
    \item Namvar, M., Intezari, A., Akhlaghpour, S. and Brienza, J.P. 2022. Beyond effective use: Integrating wise reasoning in machine learning development. \emph{International journal of information management}., 102566.
    \item Baijens, J., Helms, R. and Iren, D. 2020. Applying Scrum in Data Science Projects. \emph{Proceedings of the 22nd Conference on Business Informatics (CBI)}, 30–38.
    \item Lwakatare, L.E., Raj, A., Bosch, J., Olsson, H.H. and Crnkovic, I. 2019. A taxonomy of software engineering challenges for machine learning systems: An empirical investigation. \emph{Proceedings of the 2019 International Conference on Agile Software Development}, 227–243.
    \item Amershi, S., Begel, A., Bird, C., DeLine, R., Gall, H., Kamar, E., Nagappan, N., Nushi, B. and Zimmermann, T. 2019. Software Engineering for Machine Learning: A Case Study. \emph{Proceedings of the 41st International Conference on Software Engineering: Software Engineering in Practice (ICSE-SEIP)}, 291–300.
    \item Arpteg, A., Brinne, B., Crnkovic-Friis, L. and Bosch, J. 2018. Software Engineering Challenges of Deep Learning. \emph{Proceedings of the 44th Euromicro Conference on Software Engineering and Advanced Applications (SEAA)}, 50–59.
    \item Muiruri, D., Lwakatare, L.E., K Nurminen, J. and Mikkonen, T. 2022. Practices and Infrastructures for ML Systems--An Interview Study in Finnish Organizations. \emph{TechRxiv}.
    \item Serban, A., van der Blom, K., Hoos, H. and Visser, J. 2021. Practices for Engineering Trustworthy Machine Learning Applications. \emph{Proceedings of the 1st Workshop on AI Engineering - Software Engineering for AI (WAIN)}, 97–100.
    \item Shankar, S., Garcia, R., Hellerstein, J.M. and Parameswaran, A.G. 2022. Operationalizing Machine Learning: An Interview Study. \emph{arXiv [cs.SE]}.
    \item Andrade, H., Lwakatare, L.E., Crnkovic, I. and Bosch, J. 2019. Software Challenges in Heterogeneous Computing: A Multiple Case Study in Industry. \emph{Proceedings of the 45th Euromicro Conference on Software Engineering and Advanced Applications (SEAA)}, 148–155.
    \item Boenisch, F., Battis, V., Buchmann, N. and Poikela, M. 2021. ``I Never Thought About Securing My Machine Learning Systems'': A Study of Security and Privacy Awareness of Machine Learning Practitioners. \emph{Proceedings of Mensch und Computer 2021}, 520–546.
    \item Brennen, A. 2020. What Do People Really Want When They Say They Want ``Explainable AI?'' We Asked 60 Stakeholders. \emph{Extended Abstracts of the 2020 CHI Conference on Human Factors in Computing Systems}, 1–7.
    \item Bhatt, U., Xiang, A., Sharma, S., Weller, A., Taly, A., Jia, Y., Ghosh, J., Puri, R., Moura, J.M.F. and Eckersley, P. 2020. Explainable machine learning in deployment. \emph{Proceedings of the 2020 Conference on Fairness, Accountability, and Transparency}, 648–657.
    \item Hummer, W., Muthusamy, V., Rausch, T., Dube, P., El Maghraoui, K., Murthi, A. and Oum, P. 2019. ModelOps: Cloud-Based Lifecycle Management for Reliable and Trusted AI. \emph{Proceedings of the 2019 IEEE International Conference on Cloud Engineering (IC2E)}, 113–120.
    \item Zhang, A.X., Muller, M. and Wang, D. 2020. How do data science workers collaborate? Roles, workflows, and tools. \emph{Proceedings of the ACM on human-computer interaction}. 4, CSCW1, 1–23.
    \item Piorkowski, D., González, D., Richards, J. and Houde, S. 2020. Towards evaluating and eliciting high-quality documentation for intelligent systems. \emph{arXiv [cs.SE]}.
    \item Passi, S. and Jackson, S.J. 2018. Trust in Data Science: Collaboration, Translation, and Accountability in Corporate Data Science Projects. \emph{Proceedings of the ACM on Human-Computer Interaction}. 2, CSCW (Nov. 2018), 1–28.
    \item Dove, G., Halskov, K., Forlizzi, J. and Zimmerman, J. 2017. UX Design Innovation: Challenges for Working with Machine Learning as a Design Material. \emph{Proceedings of the 2017 CHI Conference on Human Factors in Computing Systems}, 278–288.
\end{enumerate}

\subsection*{Paper Analysis and Card Sorting}
Access the complete paper analysis and card sorting details in our online supplementary documents --

Nahar, N. 2022. Supplementary documents: A meta-summary of challenges in building products with ML components -- collecting experiences from 4758+ practitioners. OSF. \url{https://osf.io/y5edu/}

\begin{thebibliography}{100}
\bibitem{jmTM} Amershi, S., Begel, A., Bird, C., DeLine, R., Gall, H., Kamar, E., Nagappan, N., Nushi, B. and Zimmermann, T. 2019. Software Engineering for Machine Learning: A Case Study. \emph{Proceedings of the 41st International Conference on Software Engineering: Software Engineering in Practice (ICSE-SEIP)}, 291–300.
\bibitem{HaFJ} Andrade, H., Lwakatare, L.E., Crnkovic, I. and Bosch, J. 2019. Software Challenges in Heterogeneous Computing: A Multiple Case Study in Industry. \emph{Proceedings of the 45th Euromicro Conference on Software Engineering and Advanced Applications (SEAA)}, 148–155.
\bibitem{uoP6} Arnold, M., Piorkowski, D., Reimer, D., Richards, J., Tsay, J., Varshney, K.R., Bellamy, R.K.E., Hind, M., Houde, S., Mehta, S., Mojsilovic, A., Nair, R., Ramamurthy, K.N. and Olteanu, A. 2019. FactSheets: Increasing trust in AI services through supplier's declarations of conformity. \emph{IBM journal of research and development}. 63, 4/5, 6:1–6:13.
\bibitem{KOCv} Arpteg, A., Brinne, B., Crnkovic-Friis, L. and Bosch, J. 2018. Software Engineering Challenges of Deep Learning. \emph{Proceedings of the 44th Euromicro Conference on Software Engineering and Advanced Applications (SEAA)}, 50–59.
\bibitem{rPJR} Ashmore, R., Calinescu, R. and Paterson, C. 2022. Assuring the Machine Learning Lifecycle: Desiderata, Methods, and Challenges. \emph{ACM Computing Surveys}. 54, 5, 1–39.
\bibitem{UjDH} Baijens, J., Helms, R. and Iren, D. 2020. Applying Scrum in Data Science Projects. \emph{Proceedings of the 22nd Conference on Business Informatics (CBI)}, 30–38.
\bibitem{IjZD} Bäuerle, A., Cabrera, Á.A., Hohman, F., Maher, M., Koski, D., Suau, X., Barik, T. and Moritz, D. 2022. Symphony: Composing Interactive Interfaces for Machine Learning. \emph{Proceedings of the 2022 CHI Conference on Human Factors in Computing Systems}, 1–14.
\bibitem{bJXS} Begel, A. and Zimmermann, T. 2014. Analyze this! 145 questions for data scientists in software engineering. \emph{Proceedings of the 36th International Conference on Software Engineering}, 12–23.
\bibitem{x2RD} Bernardi, L., Mavridis, T. and Estevez, P. 2019. 150 Successful Machine Learning Models: 6 Lessons Learned at Booking.com. \emph{Proceedings of the 25th ACM SIGKDD International Conference on Knowledge Discovery \& Data Mining - KDD '19}, 1743–1751.
\bibitem{SKJO} Bhatt, U., Xiang, A., Sharma, S., Weller, A., Taly, A., Jia, Y., Ghosh, J., Puri, R., Moura, J.M.F. and Eckersley, P. 2020. Explainable machine learning in deployment. \emph{Proceedings of the 2020 Conference on Fairness, Accountability, and Transparency}, 648–657.
\bibitem{VlSL} Boenisch, F., Battis, V., Buchmann, N. and Poikela, M. 2021. ``I Never Thought About Securing My Machine Learning Systems'': A Study of Security and Privacy Awareness of Machine Learning Practitioners. \emph{Proceedings of Mensch und Computer 2021}, 520–546.
\bibitem{ERdQ} Borch, C. 2022. Machine learning, knowledge risk, and principal-agent problems in automated trading. \emph{Technology in society}. 68,, 101852.
\bibitem{9HsR} Borg, M., Englund, C., Wnuk, K., Duran, B., Levandowski, C., Gao, S., Tan, Y., Kaijser, H., Lönn, H. and Törnqvist, J. 2020. Safely Entering the Deep: A Review of Verification and Validation for Machine Learning and a Challenge Elicitation in the Automotive Industry. \emph{Journal of Automotive Software Engineering}. 1, 1, 1–19.
\bibitem{uF5W} Boyd, K.L. 2021. Datasheets for Datasets help ML Engineers Notice and Understand Ethical Issues in Training Data. \emph{Proceedings of the ACM on Human-Computer Interaction}. 5, CSCW2, 1–27.
\bibitem{bQQ0} Braiek, H.B. and Khomh, F. 2020. On testing machine learning programs. \emph{The Journal of systems and software}. 164,, 110542.
\bibitem{nLVo} Breck, E., Cai, S., Nielsen, E., Salib, M. and Sculley, D. 2017. The ML test score: A rubric for ML production readiness and technical debt reduction. \emph{Proceedings of the 2017 IEEE International Conference on Big Data (Big Data)}, 1123–1132.
\bibitem{a19O} Brennen, A. 2020. What Do People Really Want When They Say They Want ``Explainable AI?'' We Asked 60 Stakeholders. \emph{Extended Abstracts of the 2020 CHI Conference on Human Factors in Computing Systems}, 1–7.
\bibitem{cupz} Chang, J. and Custis, C. 2022. Understanding Implementation Challenges in Machine Learning Documentation. \emph{Equity and Access in Algorithms, Mechanisms, and Optimization}, 1–8.
\bibitem{66i9} Chattopadhyay, S., Prasad, I., Henley, A.Z., Sarma, A. and Barik, T. 2020. What's wrong with computational notebooks? Pain points, needs, and design opportunities. \emph{Proceedings of the 2020 CHI Conference on Human Factors in Computing Systems}, 1–12.
\bibitem{pV6m} Chotisarn, N., Merino, L., Zheng, X., Lonapalawong, S., Zhang, T., Xu, M. and Chen, W. 2020. A systematic literature review of modern software visualization. \emph{Journal of visualization / the Visualization Society of Japan}. 23, 4, 539–558.
\bibitem{vsJd} Dilhara, M., Ketkar, A. and Dig, D. 2021. Understanding Software-2.0: A Study of Machine Learning Library Usage and Evolution. \emph{ACM Transactions on Software Engineering and Methodology}. 30, 4, 1–42.
\bibitem{rk7b} Dove, G., Halskov, K., Forlizzi, J. and Zimmerman, J. 2017. UX Design Innovation: Challenges for Working with Machine Learning as a Design Material. \emph{Proceedings of the 2017 CHI Conference on Human Factors in Computing Systems}, 278–288.
\bibitem{qOwU} Epperson, W., Wang, A.Y., DeLine, R. and Drucker, S.M. 2022. Strategies for Reuse and Sharing among Data Scientists in Software Teams. \emph{Proceedings of the 44th International Conference on Software Engineering: Software Engineering in Practice}, 243–252.
\bibitem{UxiN} Faan, M.S.P. and Aprn, J.B.P. 2006. \emph{Handbook for Synthesizing Qualitative Research}. Springer Publishing Company.
\bibitem{ffiu} Felizardo, K.R., Mendes, E., Kalinowski, M., Souza, É.F. and Vijaykumar, N.L. 2016. Using Forward Snowballing to update Systematic Reviews in Software Engineering. \emph{Proceedings of the 10th ACM/IEEE International Symposium on Empirical Software Engineering and Measurement}, 1–6.
\bibitem{gVYg} Follow, R. Bridging the Gap Between Data Science \& Engineer: Building High-Performance Teams.
\bibitem{1nxv} Giray, G. 2021. A Software Engineering Perspective on Engineering Machine Learning Systems: State of the Art and Challenges. \emph{Journal of Systems and Software}. 180,, 111031.
\bibitem{GZrM} Golendukhina, V., Lenarduzzi, V. and Felderer, M. 2022. What is software quality for AI engineers? Towards a thinning of the fog. \emph{Proceedings of the 1st International Conference on AI Engineering: Software Engineering for AI}, 1–9.
\bibitem{XpO1} Haakman, M., Cruz, L., Huijgens, H. and van Deursen, A. 2021. AI Lifecycle Models Need To Be Revised. An Exploratory Study in Fintech. \emph{Empirical Software Engineering}. 26, 5, 1–29.
\bibitem{Ak3f} Harris, J. 2020. Beyond the jupyter notebook: how to build data science products. \emph{Towards Data Science}.
\bibitem{cHAy} Hazelwood, K. et al. 2018. Applied Machine Learning at Facebook: A Datacenter Infrastructure Perspective. \emph{Proceedings of the 2018 IEEE International Symposium on High Performance Computer Architecture (HPCA)} (Feb. 2018), 620–629.
\bibitem{rQUY} Head, A., Hohman, F., Barik, T., Drucker, S.M. and DeLine, R. 2019. Managing messes in computational notebooks. \emph{Proceedings of the 2019 CHI Conference on Human Factors in Computing Systems - CHI '19}, 1–12.
\bibitem{y7Bt} Henry, K.E., Kornfield, R., Sridharan, A., Linton, R.C., Groh, C., Wang, T., Wu, A., Mutlu, B. and Saria, S. 2022. Human-machine teaming is key to AI adoption: clinicians' experiences with a deployed machine learning system. \emph{NPJ digital medicine}. 5, 1, 1–6.
\bibitem{rN8P} Hermann, J. and Del Balso, M. 2017. Meet Michelangelo: Uber's machine learning platform.
\bibitem{ZrtG} Hill, C., Bellamy, R., Erickson, T. and Burnett, M. 2016. Trials and tribulations of developers of intelligent systems: A field study. \emph{Proceedings of the IEEE Symposium on Visual Languages and Human-Centric Computing (VL/HCC)}, 162–170.
\bibitem{O0Hy} Holstein, K., Wortman Vaughan, J., Daumé, H., Dudik, M. and Wallach, H. 2019. Improving Fairness in Machine Learning Systems: What Do Industry Practitioners Need? \emph{Proceedings of the 2019 CHI Conference on Human Factors in Computing Systems}, 1–16.
\bibitem{zo5R} Hopkins, A. and Booth, S. 2021. Machine Learning Practices Outside Big Tech: How Resource Constraints Challenge Responsible Development. \emph{Proceedings of the 2021 AAAI/ACM Conference on AI, Ethics, and Society}, 134–145.
\bibitem{hW2b} Huang, X., Kroening, D., Ruan, W., Sharp, J., Sun, Y., Thamo, E., Wu, M. and Yi, X. 2020. A survey of safety and trustworthiness of deep neural networks: Verification, testing, adversarial attack and defence, and interpretability. \emph{Computer Science Review}. 37,, 100270.
\bibitem{h6oJ} Huang, X., Zhang, H., Zhou, X., Babar, M.A. and Yang, S. 2018. Synthesizing qualitative research in software engineering: a critical review. \emph{Proceedings of the 40th International Conference on Software Engineering}, 1207–1218.
\bibitem{9tGL} Hudson, W. 2013. Card Sorting. \emph{The Encyclopedia of Human-Computer Interaction, 2nd Ed.} The Interaction Design Foundation.
\bibitem{ZAdD} Hulten, G. 2018. \emph{Building Intelligent Systems: A Guide to Machine Learning Engineering}. Apress.
\bibitem{VItT} Hummer, W., Muthusamy, V., Rausch, T., Dube, P., El Maghraoui, K., Murthi, A. and Oum, P. 2019. ModelOps: Cloud-Based Lifecycle Management for Reliable and Trusted AI. \emph{Proceedings of the 2019 IEEE International Conference on Cloud Engineering (IC2E)}, 113–120.
\bibitem{xzTz} Hynes, N., Sculley, D. and Terry, M. 2017. The data linter: Lightweight, automated sanity checking for ml data sets. \emph{NIPS MLSys Workshop}. 1,, 5.
\bibitem{QNQY} Ishikawa, F. and Yoshioka, N. 2019. How do engineers perceive difficulties in engineering of machine-learning systems? - questionnaire survey. \emph{Proceedings of the 2019 IEEE/ACM Joint 7th International Workshop on Conducting Empirical Studies in Industry (CESI) and 6th International Workshop on Software Engineering Research and Industrial Practice (SER\&IP)}, 2–9.
\bibitem{9igj} Jain, R. and Suman, U. 2015. A Systematic Literature Review on Global Software Development Life Cycle. \emph{SIGSOFT Softw. Eng. Notes}. 40, 2, 1–14.
\bibitem{b6IF} Jentzsch, S. and Hochgeschwender, N. 2021. A qualitative study of Machine Learning practices and engineering challenges in Earth Observation. \emph{it - Information Technology}. 63, 4, 235–247.
\bibitem{qky1} John, M.M., Olsson, H.H. and Bosch, J. 2020. AI Deployment Architecture: Multi-Case Study for Key Factor Identification. \emph{Proceedings of the 27th Asia-Pacific Software Engineering Conference (APSEC)}, 395–404.
\bibitem{TcaQ} Kanagal, B. and Tata, S. 2018. Recommendations for All: Solving Thousands of Recommendation Problems Daily. \emph{Proceedings of the 34th International Conference on Data Engineering (ICDE)}, 1404–1413.
\bibitem{y46l} Kästner, C. 2022. \emph{Machine Learning in Production: From Models to Products}.
\bibitem{XMGU} Keele, S. 2007. \emph{Guidelines for performing systematic literature reviews in software engineering}. Technical Rep., Ver. 2.3 EBSE Tech. Report. EBSE.
\bibitem{TtFP} Kim, M., Zimmermann, T., DeLine, R. and Begel, A. 2018. Data Scientists in Software Teams: State of the Art and Challenges. \emph{IEEE Transactions on Software Engineering}. 44, 11, 1024–1038.
\bibitem{kGqp} Kim, M., Zimmermann, T., DeLine, R. and Begel, A. 2016. The emerging role of data scientists on software development teams. \emph{Proceedings of the 38th International Conference on Software Engineering}, 96–107.
\bibitem{SCCw} Königstorfer, F. and Thalmann, S. 2022. AI Documentation: A path to accountability. \emph{Journal of Responsible Technology}. 11,, 100043.
\bibitem{omzN} Kumar, R.S.S., Nystrom, M., Lambert, J., Marshall, A., Goertzel, M., Comissoneru, A., Swann, M. and Xia, S. 2020. Adversarial Machine Learning - Industry Perspectives. \emph{Proceedings of the 2020 IEEE Security and Privacy Workshops (SPW).}, 69–75.
\bibitem{2f5l} Laato, S., Birkstedt, T., Mäantymäki, M., Minkkinen, M. and Mikkonen, T. 2022. AI governance in the system development life cycle: insights on responsible machine learning engineering. \emph{Proceedings of the 1st International Conference on AI Engineering: Software Engineering for AI}, 113–123.
\bibitem{U3jM} Lakshmanan, V., Robinson, S. and Munn, M. 2020. \emph{Machine Learning Design Patterns}. O'Reilly Media, Inc.
\bibitem{R4n4} Lewis, G.A., Bellomo, S. and Ozkaya, I. 2021. Characterizing and Detecting Mismatch in Machine-Learning-Enabled Systems. \emph{Proceedings of the IEEE/ACM 1st Workshop on AI Engineering-Software Engineering for AI (WAIN)}, 133–140.
\bibitem{5GnQ} Lewis, G.A., Ozkaya, I. and Xu, X. 2021. Software Architecture Challenges for ML Systems. \emph{Proceedings of the 2021 IEEE International Conference on Software Maintenance and Evolution (ICSME)}, 634–638.
\bibitem{JF8I} Lin, J. and Kolcz, A. 2012. Large-scale machine learning at twitter. \emph{Proceedings of the 2012 ACM SIGMOD International Conference on Management of Data}, 793–804.
\bibitem{gqhD} Li, S., Guo, J., Lou, J.-G., Fan, M., Liu, T. and Zhang, D. 2022. Testing machine learning systems in industry: an empirical study. \emph{Proceedings of the 44th International Conference on Software Engineering: Software Engineering in Practice}, 263–272.
\bibitem{6Iam} Liu, H., Eksmo, S., Risberg, J. and Hebig, R. 2020. Emerging and Changing Tasks in the Development Process for Machine Learning Systems. \emph{Proceedings of the International Conference on Software and System Processes}, 125–134.
\bibitem{clxs} Liu, J., Boukhelifa, N. and Eagan, J.R. 2020. Understanding the Role of Alternatives in Data Analysis Practices. \emph{IEEE transactions on visualization and computer graphics}. 26, 1, 66–76.
\bibitem{LOJN} Liu, Q., Li, P., Zhao, W., Cai, W., Yu, S. and Leung, V.C.M. 2018. A Survey on Security Threats and Defensive Techniques of Machine Learning: A Data Driven View. \emph{IEEE Access}. 6,, 12103–12117.
\bibitem{3bBp} Lopez, G. and Guerrero, L.A. 2017. Awareness Supporting Technologies used in Collaborative Systems: A Systematic Literature Review. \emph{Proceedings of the 2017 ACM Conference on Computer Supported Cooperative Work and Social Computing}, 808–820.
\bibitem{x37l} Lwakatare, L.E., Raj, A., Bosch, J., Olsson, H.H. and Crnkovic, I. 2019. A taxonomy of software engineering challenges for machine learning systems: An empirical investigation. \emph{Proceedings of the 2019 International Conference on Agile Software Development}, 227–243.
\bibitem{G64S} Lwakatare, L.E., Raj, A., Crnkovic, I., Bosch, J. and Olsson, H.H. 2020. Large-scale machine learning systems in real-world industrial settings: A review of challenges and solutions. \emph{Information and software technology}. 127, 106368, 106368.
\bibitem{DhfH} Mäkinen, S., Skogström, H., Laaksonen, E. and Mikkonen, T. 2021. Who Needs MLOps: What Data Scientists Seek to Accomplish and How Can MLOps Help? \emph{Proceedings of the IEEE/ACM 1st Workshop on AI Engineering - Software Engineering for AI (WAIN)}, 109–112.
\bibitem{Jbnm} Martínez-Fernández, S., Bogner, J., Franch, X., Oriol, M., Siebert, J., Trendowicz, A., Vollmer, A.M. and Wagner, S. 2022. Software Engineering for AI-Based Systems: A Survey. \emph{ACM Transactions on Software Engineering and Methodology}. 31, 2, 1–59.
\bibitem{E5GQ} Martinez-Plumed, F., Contreras-Ochando, L., Ferri, C., Hernandez Orallo, J., Kull, M., Lachiche, N., Ramirez Quintana, M.J. and Flach, P.A. 2020. CRISP-DM twenty years later: From data mining processes to data science trajectories. \emph{IEEE transactions on knowledge and data engineering}. 33, 8, 3048–3061.
\bibitem{uJDv} McGlohon, M. 2021. Demystifying Machine Learning in Production: Reasoning about a Large-Scale ML Platform.
\bibitem{2FRT} McGraw, G., Figueroa, H., Shepardson, V. and Bonett, R. 2020. An architectural risk analysis of machine learning systems: Toward more secure machine learning. \emph{Berryville Institute of Machine Learning, Clarke County, VA. Accessed on: Mar}. 23,.
\bibitem{UMwZ} Mitchell, M., Wu, S., Zaldivar, A., Barnes, P., Vasserman, L., Hutchinson, B., Spitzer, E., Raji, I.D. and Gebru, T. 2019. Model Cards for Model Reporting. \emph{Proceedings of the Conference on Fairness, Accountability, and Transparency}, 220–229.
\bibitem{aJfF} Muiruri, D., Lwakatare, L.E., K Nurminen, J. and Mikkonen, T. 2022. Practices and Infrastructures for ML Systems--An Interview Study in Finnish Organizations. \emph{TechRxiv}.
\bibitem{jM0j} Muller, M., Lange, I., Wang, D., Piorkowski, D., Tsay, J., Liao, Q.V., Dugan, C. and Erickson, T. 2019. How Data Science Workers Work with Data: Discovery, Capture, Curation, Design, Creation. \emph{Proceedings of the 2019 CHI Conference on Human Factors in Computing Systems}, 1–15.
\bibitem{W82W} Myllyaho, L., Raatikainen, M., Männistö, T., Nurminen, J.K. and Mikkonen, T. 2022. On misbehaviour and fault tolerance in machine learning systems. \emph{Journal of Systems and Software}. 183,, 111096.
\bibitem{1JnV} Nahar, N. 2022. Supplementary documents: A meta-summary of challenges in building products with ML components -- collecting experiences from 4758+ practitioners. OSF. \url{https://osf.io/y5edu/}
\bibitem{xMO3} Nahar, N., Zhou, S., Lewis, G. and Kästner, C. 2022. Collaboration Challenges in Building ML-Enabled Systems: Communication, Documentation, Engineering, and Process. \emph{Proceedings of the 44th International Conference on Software Engineering}, 413–425.
\bibitem{XFkw} Namvar, M., Intezari, A., Akhlaghpour, S. and Brienza, J.P. 2022. Beyond effective use: Integrating wise reasoning in machine learning development. \emph{International journal of information management}., 102566.
\bibitem{u0v7} Nascimento, E., Nguyen-Duc, A., Sundbø, I. and Conte, T. 2020. Software engineering for artificial intelligence and machine learning software: A systematic literature review. \emph{arXiv [cs.SE]}.
\bibitem{AfqY} Nikanjam, A., Morovati, M.M., Khomh, F. and Ben Braiek, H. 2022. Faults in deep reinforcement learning programs: a taxonomy and a detection approach. \emph{Automated software engineering}. 29, 1.
\bibitem{R6PE} Nikhil, K., Anandayuvaraj, D., Detti, A., Lee Bland, F., Rahaman, S. and Davis, J.C. 2022. ``If security is required'': Engineering and Security Practices for Machine Learning-based IoT Devices. \emph{Proceedings of the 4th International Workshop on Software Engineering Research and Practices for the IoT (SERP4IoT)}, 1–8.
\bibitem{MYVR} Nushi, B., Kamar, E., Horvitz, E. and Kossmann, D. 2017. On human intellect and machine failures: troubleshooting integrative machine learning systems. \emph{Proceedings of the Thirty-First AAAI Conference on Artificial Intelligence}, 1017–1025.
\bibitem{xuax} Ozkaya, I. 2020. What is really different in engineering AI-enabled systems? \emph{IEEE software}. 37, 4, 3–6.
\bibitem{jAIR} Paleyes, A., Urma, R.-G. and Lawrence, N.D. 2022. Challenges in deploying machine learning: A survey of case studies. \emph{ACM computing surveys}..
\bibitem{auDR} Passi, S. and Jackson, S.J. 2018. Trust in Data Science: Collaboration, Translation, and Accountability in Corporate Data Science Projects. \emph{Proceedings of the ACM on Human-Computer Interaction}. 2, CSCW (Nov. 2018), 1–28.
\bibitem{5ezP} Passi, S. and Sengers, P. 2020. Making data science systems work. \emph{Big data \& society}. 7, 2, 205395172093960.
\bibitem{eI6i} Pimentel, J.F., Murta, L., Braganholo, V. and Freire, J. 2019. A large-scale study about quality and reproducibility of jupyter notebooks. \emph{Proceedings of the 16th International Conference on Mining Software Repositories (MSR)}, 507–517.
\bibitem{QOKm} Piorkowski, D., González, D., Richards, J. and Houde, S. 2020. Towards evaluating and eliciting high-quality documentation for intelligent systems. \emph{arXiv [cs.SE]}.
\bibitem{FLpR} Piorkowski, D., Park, S., Wang, A.Y., Wang, D., Muller, M. and Portnoy, F. 2021. How AI Developers Overcome Communication Challenges in a Multidisciplinary Team: A Case Study. \emph{Proceedings of the ACM on Human-Computer Interaction 5.CSCW1}, 1–25.
\bibitem{M56h} Polyzotis, N., Roy, S., Whang, S.E. and Zinkevich, M. 2018. Data Lifecycle Challenges in Production Machine Learning: A Survey. \emph{ACM SIGMOD Record}. 47, 2, 17–28.
\bibitem{HVfD} Rahimi, M., Guo, J.L.C., Kokaly, S. and Chechik, M. 2019. Toward Requirements Specification for Machine-Learned Components. \emph{Proceedings of the 27th International Requirements Engineering Conference Workshops (REW)}, 241–244.
\bibitem{ke7w} Rahman, M.S., Khomh, F., Hamidi, A., Cheng, J., Antoniol, G. and Washizaki, H. 2021. Machine Learning Application Development: Practitioners' Insights. \emph{arXiv [cs.SE]}.
\bibitem{JvNS} Rahman, M.S., Rivera, E., Khomh, F., Guéhéneuc, Y.-G. and Lehnert, B. 2019. Machine Learning Software Engineering in Practice: An Industrial Case Study. \emph{arXiv [cs.SE]}.
\bibitem{aqp3} Rakova, B., Yang, J., Cramer, H. and Chowdhury, R. 2020. Where Responsible AI meets Reality: Practitioner Perspectives on Enablers for shifting Organizational Practices. \emph{Proceedings of the ACM on Human-Computer Interaction}, 1–23.
\bibitem{J0kZ} Ribeiro, D.M., Cardoso, M., da Silva, F.Q.B. and França, C. 2014. Using qualitative metasummary to synthesize empirical findings in literature reviews. \emph{Proceedings of the 8th ACM/IEEE International Symposium on Empirical Software Engineering and Measurement}, 1–4.
\bibitem{wvQU} Ribeiro, M.T., Wu, T., Guestrin, C. and Singh, S. 2020. Beyond Accuracy: Behavioral Testing of NLP models with CheckList. \emph{arXiv [cs.CL]}.
\bibitem{InOi} Riccio, V., Jahangirova, G., Stocco, A., Humbatova, N., Weiss, M. and Tonella, P. 2020. Testing machine learning based systems: a systematic mapping. \emph{Empirical Software Engineering}. 25, 6, 5193–5254.
\bibitem{fCMZ} Rismani, S., Shelby, R., Smart, A., Jatho, E., Kroll, J., Moon, A. and Rostamzadeh, N. 2022. From plane crashes to algorithmic harm: applicability of safety engineering frameworks for responsible ML. \emph{arXiv [cs.HC]}.
\bibitem{g6Jy} Riungu-Kalliosaari, L., Kauppinen, M. and Männistö, T. 2017. What Can Be Learnt from Experienced Data Scientists? A Case Study. \emph{Product-Focused Software Process Improvement}, 55–70.
\bibitem{HQ7E} Saha, D., Schumann, C., McElfresh, D.C., Dickerson, J.P., Mazurek, M.L. and Tschantz, M.C. 2020. Human Comprehension of Fairness in Machine Learning. \emph{Proceedings of the AAAI/ACM Conference on AI, Ethics, and Society}, 152.
\bibitem{E34o} Salay, R., Queiroz, R. and Czarnecki, K. 2017. An Analysis of ISO 26262: Using Machine Learning Safely in Automotive Software. \emph{arXiv [cs.AI]}.
\bibitem{onib} Sambasivan, N., Kapania, S., Highfill, H., Akrong, D., Paritosh, P. and Aroyo, L.M. 2021. ``Everyone wants to do the model work, not the data work'': Data Cascades in High-Stakes AI. \emph{Proceedings of the 2021 CHI Conference on Human Factors in Computing Systems}, 1–15.
\bibitem{ZbHh} Sandelowski, M., Barroso, J. and Voils, C.I. 2007. Using qualitative metasummary to synthesize qualitative and quantitative descriptive findings. \emph{Research in nursing \& health}. 30, 1, 99–111.
\bibitem{RlvD} Sculley, D., Holt, G., Golovin, D., Davydov, E., Phillips, T., Ebner, D., Chaudhary, V., Young, M., Crespo, J.-F. and Dennison, D. 2015. Hidden Technical Debt in Machine Learning Systems. \emph{Advances in Neural Information Processing Systems 28}. C. Cortes, N.D. Lawrence, D.D. Lee, M. Sugiyama, and R. Garnett, eds. Curran Associates, Inc. 2503–2511.
\bibitem{JQhY} Sculley, D., Otey, M.E., Pohl, M., Spitznagel, B., Hainsworth, J. and Zhou, Y. 2011. Detecting adversarial advertisements in the wild. \emph{Proceedings of the 17th ACM SIGKDD international conference on Knowledge discovery and data mining}, 274–282.
\bibitem{FBN2} Sendak, M.P. et al. 2020. Real-World Integration of a Sepsis Deep Learning Technology Into Routine Clinical Care: Implementation Study. \emph{JMIR medical informatics}. 8, 7, e15182.
\bibitem{wHQl} Serban, A., van der Blom, K., Hoos, H. and Visser, J. 2020. Adoption and Effects of Software Engineering Best Practices in Machine Learning. \emph{Proceedings of the 14th ACM/IEEE International Symposium on Empirical Software Engineering and Measurement (ESEM)}, 1–12.
\bibitem{7fVO} Serban, A., van der Blom, K., Hoos, H. and Visser, J. 2021. Practices for Engineering Trustworthy Machine Learning Applications. \emph{Proceedings of the 1st Workshop on AI Engineering - Software Engineering for AI (WAIN)}, 97–100.
\bibitem{Dk4s} Serban, A. and Visser, J. 2022. Adapting Software Architectures to Machine Learning Challenges. \emph{Proceedings of the 2022 IEEE International Conference on Software Analysis, Evolution and Reengineering (SANER)}, 152–163.
\bibitem{4myf} Shankar, S., Garcia, R., Hellerstein, J.M. and Parameswaran, A.G. 2022. Operationalizing Machine Learning: An Interview Study. \emph{arXiv [cs.SE]}.
\bibitem{8qrd} Shaw, M. and Zhu, L. 2022. Can Software Engineering Harness the Benefits of Advanced AI? \emph{IEEE Software}. 39, 6, 99–104.
\bibitem{AI7y} Siebert, J., Joeckel, L., Heidrich, J., Nakamichi, K., Ohashi, K., Namba, I., Yamamoto, R. and Aoyama, M. 2020. Towards Guidelines for Assessing Qualities of Machine Learning Systems. \emph{Proceedings of the International Conference on the Quality of Information and Communications Technology}, 17–31.
\bibitem{Brt4} Smith, D. 2017. Exploring development patterns in data science.
\bibitem{MTwb} d. S. Nascimento, E., Ahmed, I., Oliveira, E., Palheta, M.P., Steinmacher, I. and Conte, T. 2019. Understanding Development Process of Machine Learning Systems: Challenges and Solutions. \emph{Proceedings of the 2019 ACM/IEEE International Symposium on Empirical Software Engineering and Measurement (ESEM)}, 1–6.
\bibitem{bBfc} Spencer, D. 2009. \emph{Card Sorting: Designing Usable Categories}. Rosenfeld Media.
\bibitem{Hi9f} Studer, S., Bui, T.B., Drescher, C., Hanuschkin, A., Winkler, L., Peters, S. and Mueller, K.-R. 2021. Towards CRISP-ML(Q): A Machine Learning Process Model with Quality Assurance Methodology. \emph{Machine Learning and Knowledge Extraction}. 3, 2, 392–413.
\bibitem{3md1} Tonekaboni, S., Joshi, S., McCradden, M.D. and Goldenberg, A. 2019. What Clinicians Want: Contextualizing Explainable Machine Learning for Clinical End Use. \emph{Proceedings of the 4th Machine Learning for Healthcare Conference}, 359–380.
\bibitem{kw6g} Uchihira, N. 2022. Project FMEA for Recognizing Difficulties in Machine Learning Application System Development. \emph{Proceedings of the 2022 Portland International Conference on Management of Engineering and Technology (PICMET)}, 1–8.
\bibitem{DNDN} Vogelsang, A. and Borg, M. 2019. Requirements Engineering for Machine Learning: Perspectives from Data Scientists. \emph{Proceedings of the 27th International Requirements Engineering Conference Workshops (REW)}, 245–251.
\bibitem{75w3} Wagstaff, K. 2012. Machine Learning that Matters. \emph{arXiv [cs.LG]}.
\bibitem{agH0} Wang, D., Weisz, J.D., Muller, M., Ram, P., Geyer, W., Dugan, C., Tausczik, Y., Samulowitz, H. and Gray, A. 2019. Human-AI Collaboration in Data Science: Exploring Data Scientists' Perceptions of Automated AI. \emph{Proceedings of the ACM on Human-Computer Interaction}. 3, CSCW, 1–24.
\bibitem{IrZw} Wan, Z., Xia, X., Lo, D. and Murphy, G.C. 2019. How does Machine Learning Change Software Development Practices? \emph{IEEE Transactions on Software Engineering}. 47, 9, 1857–1871.
\bibitem{AaNn} Washizaki, H., Takeuchi, H., Khomh, F., Natori, N., Doi, T. and Okuda, S. 2020. Practitioners' insights on machine-learning software engineering design patterns: a preliminary study. \emph{Proceedings of the 2020 IEEE International Conference on Software Maintenance and Evolution (ICSME)}, 797–799.
\bibitem{YY1S} Washizaki, H., Uchida, H., Khomh, F. and Guéhéneuc, Y.-G. 2020. Machine learning architecture and design patterns. \emph{IEEE Software}. 8,.
\bibitem{eAi9} Washizaki, H., Uchida, H., Khomh, F. and Guéhéneuc, Y.-G. 2019. Studying Software Engineering Patterns for Designing Machine Learning Systems. \emph{Proceedings of the 10th International Workshop on Empirical Software Engineering in Practice (IWESEP)}, 49–495.
\bibitem{oFbx} Wohlin, C. 2014. Guidelines for snowballing in systematic literature studies and a replication in software engineering. \emph{Proceedings of the 18th International Conference on Evaluation and Assessment in Software Engineering}, 1–10.
\bibitem{Yy9q} Zdanowska, S. and Taylor, A.S. 2022. A study of UX practitioners roles in designing real-world, enterprise ML systems. \emph{Proceedings of the 2022 CHI Conference on Human Factors in Computing Systems}, 1–15.
\bibitem{DKy1} Zhang, A.X., Muller, M. and Wang, D. 2020. How do data science workers collaborate? Roles, workflows, and tools. \emph{Proceedings of the ACM on human-computer interaction}. 4, CSCW1, 1–23.
\bibitem{PSLS} Zhang, J.M., Harman, M., Ma, L. and Liu, Y. 2022. Machine learning testing: Survey, landscapes and horizons. \emph{IEEE Transactions on Software Engineering}. 48, 1, 1–36.
\bibitem{CN5W} Zhang, X., Yang, Y., Feng, Y. and Chen, Z. 2019. Software Engineering Practice in the Development of Deep Learning Applications. \emph{arXiv [cs.SE]}.
\bibitem{0BTr} Zinkevich, M. 2017. Rules of machine learning: Best practices for ML engineering. minegrado.ovh.
\end{thebibliography}
\end{document}